\documentclass[reqno,11pt]{amsart}
\usepackage{graphicx}
\usepackage{amscd}
\usepackage{tensor}
\usepackage{amssymb}
\usepackage{esint}
\usepackage{enumitem}
\usepackage{accents}
\usepackage{pst-grad} 
\usepackage{pst-plot} 
\usepackage[mathscr]{eucal}
\numberwithin{equation}{section}
\allowdisplaybreaks[1]

\title[The Fermionic Signature Operator in the Schwarzschild Geometry]
{The Fermionic Signature Operator in the \\ Exterior Schwarzschild Geometry}

\author[F.\ Finster]{Felix Finster}
\address{Fakult\"at f\"ur Mathematik \\ Universit\"at Regensburg \\ D-93040 Regensburg \\ Germany}\email{finster@ur.de}
\author[C.\ R\"oken]{Christian R\"oken \\ \\ December 2018}
\address{Departamento de Geometr\'ia y Topolog\'ia, Facultad de Ciencias - Universidad de Gra\-na\-da, Campus de Fuentenueva s/n, 18071 Granada, Spain}
\thanks{Supported by the DFG research grant ``Dirac Waves in the Kerr Geometry: Integral Representations, Mass Oscillation Property and the Hawking Effect.''}
\email{Christian.Roeken@gmx.de}

\newtheorem{Def}{Definition}[section]
\newtheorem{Thm}[Def]{Theorem}

\newtheorem{Lemma}[Def]{Lemma}

\newtheorem{Corollary}[Def]{Corollary}

\newcommand{\Thanks}{\vspace*{.5em} \noindent \thanks}
\newcommand{\beq}{\begin{equation}}
\newcommand{\eeq}{\end{equation}}
\newcommand{\Proof}{\begin{proof}}
\newcommand{\QED}{\end{proof} \noindent}

\newcommand{\la}{\langle}
\newcommand{\ra}{\rangle}
\newcommand{\bra}{\mathopen{<}}
\newcommand{\ket}{\mathclose{>}}
\newcommand{\Sl}{\mbox{$\prec \!\!$ \nolinebreak}}
\newcommand{\Sr}{\mbox{\nolinebreak $\succ$}}

\newcommand{\C}{\mathbb{C}}
\newcommand{\R}{\mathbb{R}}
\newcommand{\1}{\mbox{\rm 1 \hspace{-1.05 em} 1}}
\newcommand{\Z}{\mathbb{Z}}

\renewcommand{\O}{{\mathscr{O}}}

\newcommand\B{{\mathscr{B}}}

\newcommand{\Cisc}{C^\infty_{\text{\rm{sc}}}}
\newcommand{\Cisco}{C^\infty_{\text{\rm{sc}},0}}
\newcommand{\Dir}{{\mathcal{D}}}
\DeclareMathOperator{\supp}{supp}
\renewcommand{\H}{\mathscr{H}}

\newcommand{\Lin}{\text{\rm{L}}}

\newcommand{\D}{\mathscr{D}}

\newcommand{\p}{\mathfrak{p}}

\newcommand{\Sig}{\mathscr{S}}

\newcommand{\scrM}{\mycal M}
\newcommand{\scrN}{\mycal N}

\DeclareFontFamily{OT1}{rsfso}{}
\DeclareFontShape{OT1}{rsfso}{m}{n}{ <-7> rsfso5 <7-10> rsfso7 <10-> rsfso10}{}
\DeclareMathAlphabet{\mycal}{OT1}{rsfso}{m}{n}

\newcommand{\bitem}{\begin{itemize}[leftmargin=2em]}
\newcommand{\eitem}{\end{itemize}}

\begin{document}

\begin{abstract}
The structure of the solution space of the Dirac equation in the exterior
Schwarzschild geometry is analyzed.
Representing the space-time inner product for families of solutions
with variable mass parameter in terms of the respective scalar products,
a so-called mass decomposition is derived.
This mass decomposition consists of a single mass integral involving the fermionic signature operator
as well as a double integral which takes into account the flux of Dirac currents across the event horizon.
The spectrum of the fermionic signature operator is computed.
The corresponding generalized fermionic projector states are
analyzed.
\end{abstract}

\maketitle
\tableofcontents

\section{Introduction}
The fermionic signature operator introduced in~\cite{finite, infinite}
gives a general setting for spectral geometry in Lorentzian signature~\cite{drum}
and is useful for constructing quasi-free Dirac states
in globally hyperbolic space-times~\cite{fewster+lang, hadamard}.
In the present paper, the fermionic signature operator is constructed for the first time in
a {\em{black hole geometry}}, namely the exterior Schwarzschild geometry.
 The event horizon makes it necessary to modify the
constructions considerably. In order to explain these modifications, we briefly recall
the general idea and the basic construction in~\cite{finite, infinite}
(the necessary preliminaries will be provided in Section~\ref{secprelim} below).
On solutions~$\psi_m, \phi_m$ of the Dirac equation of mass~$m$
in a globally hyperbolic space-time~$(\scrM, g)$, one has two
inner products: One is the scalar product obtained by integrating the polarized probability density over a
Cauchy surface~$\scrN$,
\beq (\psi_m | \phi_m)_m := 2 \pi \int_\scrN \Sl \psi_m \,|\, \gamma^j \nu_j\, \phi_m \Sr_x\: d\mu_\scrN(x) \label{printintro}
\eeq
(where~$\nu$ is the future-directed normal), whereas the other is
obtained by integrating the pointwise inner product of Dirac wave functions over all of space-time,
\beq
\bra \psi|\phi \ket := \int_\scrM \Sl \psi | \phi \Sr_x \: d\mu_\scrM \label{stipintro} \:.
\eeq
The scalar product~\eqref{printintro} endows the solution space of
the Dirac equation with the structure of a Hilbert space~$(\H_m, (.|.)_m)$.
In non-technical terms, the fermionic signature operator arises when representing
the space-time inner product~$\bra .|. \ket$ with respect to the scalar product~$(.|.)_m$.
In space-times of {\em{finite lifetime}}~\cite{finite}, this method can be implemented directly
by demanding that the relation
\beq \label{rep}
\bra \psi_m | \phi_m \ket = (\psi_m | \Sig_m \phi_m)_m
\eeq
should hold for all~$\psi_m, \phi_m \in \H_m$. This uniquely defines~$\Sig_m$ as a symmetric
bounded operator on~$\H_m$.
In space-times of {\em{infinite lifetime}}~\cite{infinite}, the relation~\eqref{rep} 
in general is not sensible (except in specific situations like the Rindler space-time~\cite{rindler}),
simply because the time integration in~\eqref{stipintro} may diverge for Dirac solutions.
The way out is to make use of {\em{mass oscillations}} in the following sense.
Instead of analyzing solutions for a fixed mass~$m$, one considers
families~$(\psi_m)_{m \in I}$ of Dirac solutions for a mass parameter~$m$
which varies in an interval~$I := (m_L, m_R)$ with~$0 \not\in I$. 
Integrating over the mass parameter,
\[ \p \psi := \int_I \psi_m\: dm \:, \]
we obtain a superposition of waves oscillating with different
frequencies. Intuitively speaking, this leads to destructive interference,
giving rise to the desired decay of the Dirac wave functions for large times.
This makes it possible to replace~\eqref{rep} by the condition
\beq \label{deltamrep}
\bra \p \psi | \p \phi \ket = \int_I (\psi_m \,|\, \Sig_m \,\phi_m)_m\: dm \:,
\eeq
to be satisfied for all families of solutions~$(\psi_m)_{m \in I}$ and~$(\phi_m)_{m \in I}$
which lie in a suitably chosen dense subspace~$\H^\infty \subset \H$ of the Hilbert space
of families of solutions (for details see Section~\ref{secdirglobhyp}). 
The space~$\H^\infty$ is referred to as the {\em{domain}} for the mass oscillations.
This construction gives for every~$m \in I$ a uniquely defined bounded linear operator~$\Sig_m$ on~$\H_m$.
The conditions needed for the construction to work are subsumed in various
notions of {\em{mass oscillation properties}}. For details we refer the interested reader to the general construction
in~\cite{infinite} and to the applications in~\cite{hadamard, planewave}.

In order to put our results into context, let us say a few general words on the fermionic signature
operator and its significance. The fermionic signature operator~$\Sig_m$ is a symmetric operator
on the Hilbert space~$\H_m$. Since its construction involves the Dirac wave functions,
it clearly contains information on the solutions of the Dirac equation.
However, since only the inner products~$(.|.)_m$ and~$\bra .|. \ket$ of these wave functions are used,
the Dirac wave functions do not enter pointwise, but merely integrated over spacd and space-time.
The construction is covariant and does not depend on observers nor on the choice of Cauchy
surfaces. Moreover, as is made precise in~\cite{sigsymm}, the fermionic signature operator respects
all space-time symmetries. To summarize, the fermionic signature operator
encodes specific information on the global behavior of the Dirac wave functions in space-time.
Apart from being of independent interest as a novel geometric operator, the fermionic signature operator
has two main applications. The first application is that it provides a setting for spectral geometry
with Lorentzian signature. This has been explored for two-dimensional space-times in~\cite{drum}.
As the second application, it gives a new method for constructing quasi-free states of the quantized Dirac field.
This is based on the observation that spectral subspaces of~$\Sig_m$ are distinguished subspaces of~$\H_m$
characterized purely geometrically independent of observers.
These distinguished subspaces can be used to construct distinguished quasi-free Dirac states,
referred to as the {\em{fermionic projector state}} or generalized fermionic projector states
(see~\cite{finite, fewster+lang} and~\cite{infinite, planewave, rindler, hadamard, desitter}).

In an exterior black hole geometry, the main complication is that part of the Dirac wave
may cross the event horizon and disappear in the black hole. 
As a consequence, the mass oscillation properties
no longer hold, and a representation of the form~\eqref{deltamrep} no longer exists.
Instead, based on the integral representation of the Dirac propagator~\cite{tkerr},
we derive a so-called {\em{mass decomposition}} of the form (for details see Theorem~\ref{thmstiprep})
\begin{align}
\bra \p \psi \,|\, \p \phi \ket 
&= \int_I (\psi_m \,|\, \Sig_m \,\phi_m)_m\: dm \label{massdecomp1} \\
&\quad\:+ \frac{i}{\pi} \int_I dm \int_I dm' \: \frac{\text{\rm{PP}}}{m-m'} \: {\mathfrak{B}}(\psi_m, \phi_{m'}) \:, \label{massdecomp2}
\end{align}
where~${\mathfrak{B}}(\psi_m, \phi_{m'})$ is a smooth function in~$m$ and~$m'$
(and~$\text{PP}$ denotes the principal value). The above equation holds for all
families~$(\psi_m)_{m \in I}$ and~$(\phi_m)_{m \in I}$
in a conveniently chosen domain~$\H^\infty$ (for details see Definition~\ref{defHinf}).
We point out that~\eqref{massdecomp2}
gives a contribution for pairs of solutions~$\psi_m$ and~$\phi_{m'}$
of the Dirac equation with different masses~$m \neq m'$.
This contribution can be associated to the flux of a corresponding
``current'' $J^k(x) = \Sl \psi_m | \gamma^j \phi_{m'}\Sr_x$ through the event horizon
of the black hole (this connection is explained in Section~\ref{secinterpret}
and worked out in Section~\ref{secB} using the so-called {\em{fermionic flux operator}}).

The contribution~\eqref{massdecomp1} to the mass decomposition again uniquely defines
for every $m \in I$ a {\em{fermionic signature operator}}~$\Sig_m$.
We analyze the properties of this operator. Our results are summarized as follows.
First of all, the fermionic signature operator is a bounded selfadjoint operator on~$\H_m$
with~$\|\Sig_m\| \leq 2$. It respects the symmetries of space-time, meaning that
it has a joint spectral decomposition with the angular momentum operator~${\mathcal{A}}$ and the
Dirac Hamiltonian~$H$ of the form
\beq \label{Sigrep}
\Sig_m = \sum_{k,n} \int_{-\infty}^\infty \Sig^{kn}(\omega)\: F_{k,n} \: dE_\omega \:,
\eeq
where~$F$ and~$E$ are the spectral measures of the operators~${\mathcal{A}}$ and~$H$, 
respectively; i.e.
\beq \label{Edef}
{\mathcal{A}} = \sum_{k,n} \lambda_n \,F_{k,n} \qquad \text{and} \qquad
H = \int_{-\infty}^\infty \omega\: dE_\omega
\eeq
(here~$k \in \Z + \frac{1}{2}$ is the azimuthal eigenvalue, and~$n \in \Z$ labels
the eigenvalues of the spin-weighted angular operator~${\mathcal{A}}$; for details
see Section~\ref{secsep} below).
For clarity, we note that the spectral decomposition~\eqref{Sigrep} already follows
abstractly from the fact that the space-time symmetries can be described
by local groups of isomorphisms of the spinor bundle~\cite{sigsymm}.
Here we obtain this representation with a computational approach, which has the great advantage that
we get detailed information on the eigenvalues of the operators~$\Sig^{kn}(\omega)$
in~\eqref{Sigrep} (for details see Theorem~\ref{thmsig}):
\bitem
\item[(i)] The operators~$\Sig^{kn}(\omega)$ vanish if~$\omega \in [-m,m]$.
\item[(ii)] In the range~$\omega \in \R \setminus [-m,m]$, the operator
\beq \label{Sigsign}
\text{$\Sig^{kn}(\omega)$ is } \left\{ \begin{array}{cl} \text{positive definite} &\text{if~$\omega>m$ } \\
\text{negative definite} &\text{if~$\omega<-m$} \:. \end{array} \right.
\eeq
Its eigenvalues are given by
\beq \label{mupm}
\mu_\pm(\omega) = \epsilon(\omega) \pm \sqrt{ \frac{\big\|f^{k\omega n}_{\infty,\:m,1} \big\|^2_{\C^2} - 1}
{\big\|f^{k\omega n}_{\infty,\:m,1} \big\|^2_{\C^2} +1}} \:,
\eeq
where~$\epsilon$ is the sign function and~$f^{k\omega n}_{\infty,\:m,1}$ are the transmission coefficients of the radial ODE (for details see Section~\ref{secfs}).
\eitem
These results show that the fermionic signature operator contains
surprisingly rich information on the black hole geometry and on
properties of the Dirac solutions: According to~(i), the kernel of~$\Sig_m$
consists of all Dirac solutions which necessarily ``fall into'' the black hole because
their kinetic energy is not large enough for the wave to propagate to the asymptotic end.
According to~\eqref{Sigsign}, the positive and negative spectral subspaces
of~$\Sig_m$ yield the frequency splitting for an observer in a rest frame at
infinity. Finally, the formula~\eqref{mupm} shows that
the gravitational force acting on the Dirac wave functions has an interesting
influence on the spectrum of~$\Sig_m$.

We also analyze the corresponding  {\em{fermionic projector state}}.
It is obtained by applying Araki's construction in~\cite{araki1970quasifree}
to the projection operator onto the negative spectral subspace
of the fermionic signature operator (for details see~\cite[Section~6]{hadamard}).
In view of~\eqref{Sigsign}, we obtain the following result:
\begin{Corollary} \label{corFRV1}
The pure quasi-free fermionic projector state obtained from the fermio\-nic signature operator
coincides with the Hadamard state which is obtained by frequency splitting for the
observer in a rest frame at infinity.
\end{Corollary} \noindent
Having non-trivial eigenvalues~\eqref{mupm}, one obtains many other
quasi-free states by applying Araki's construction to the
positive operators~$W(\Sig_m)$ with~$W$ a non-negative Borel function.
However, at present the physical significance of these so-called
{\em{generalized fermionic projector states}} is unclear.
These states are in general not Hadamard (for details see Section~\ref{secFP}).

The paper is organized as follows.
In Section~\ref{secprelim} we give the necessary background on the
Dirac equation in globally hyperbolic space-times and in the exterior Schwarzschild
geometry. The main point is to specialize the integral representation
of the Dirac operator in the Kerr geometry
which was derived and analyzed in~\cite{kerr, tkerr}
to the exterior Schwarzschild geometry. We closely follow the procedure in these
papers and use a similar notation. In Section~\ref{secmassdecomp} 
the mass decomposition~\eqref{massdecomp1} and~\eqref{massdecomp2} is derived.
In Section~\ref{secSig} the fermionic signature operator~$\Sig_m$ is computed and analyzed.
In Section~\ref{secB} we define and analyze the fermionic flux operator~$\B_m$ which
describes the flux of Dirac currents through the event horizon.
Finally, Section~\ref{secFP} is devoted to the resulting quasi-free quantum states.

\section{Preliminaries} \label{secprelim}
\subsection{The Dirac Equation in Globally Hyperbolic Space-Times} \label{secdirglobhyp}
We recall the setting in~\cite{finite, infinite}, restricting attention to four-dimensional space-times.
Thus we let~$(\scrM, g)$ be a smooth, globally hyperbolic Lorentzian spin
manifold of dimension four. For the signature of the metric we use the convention~$(+ ,-, -, -)$.
We denote the corresponding spinor bundle by~$S\scrM$. Its fibres~$S_x\scrM$ are endowed
with an inner product~$\Sl .|. \Sr_x$ of signature~$(2,2)$, referred to as the spin scalar product.
Clifford multiplication is described by a mapping~$\gamma$
which satisfies the anti-commutation relations,
\[ \gamma \::\: T_x\scrM \rightarrow \Lin(S_x\scrM) \qquad
\text{with} \qquad \gamma(u) \,\gamma(v) + \gamma(v) \,\gamma(u) = 2 \, g(u,v)\,\1_{S_x(\scrM)} \:. \]
We write Clifford multiplication in components with the Dirac matrices~$\gamma^j$.
The metric connections on the tangent bundle and the spinor bundle are denoted by~$\nabla$.
The sections of the spinor bundle are also referred to as wave functions.
We denote the smooth sections of the spinor bundle by~$C^\infty(\scrM, S\scrM)$.
Similarly, $C^\infty_0(\scrM, S\scrM)$ denotes the smooth sections with compact support.
On the wave functions, one has the Lorentz invariant inner product
\begin{gather}
\bra .|. \ket \::\: C^\infty(\scrM, S\scrM) \times C^\infty_0(\scrM, S\scrM) \rightarrow \C \:, \notag \\
\bra \psi|\phi \ket = \int_\scrM \Sl \psi | \phi \Sr_x \: d\mu_\scrM\:.  \label{stip} 
\end{gather}

The Dirac operator~$\Dir$ in a gravitational field is defined by
\[ \Dir := i \gamma^j \nabla_j \::\: C^\infty(\scrM, S\scrM) \rightarrow C^\infty(\scrM, S\scrM)\:. \]
For a given real parameter~$m \in \R$ (the ``mass''), the Dirac equation reads
\[ 
(\Dir - m) \,\psi_m = 0 \:. \]
For clarity, we always denote solutions of the Dirac equation by a subscript~$m$.
The assumption of global hyperbolicity yields the existence of a smooth foliation by
Cauchy surfaces. Given smooth initial data on a Cauchy surface~$\scrN$,
the Dirac equation has a unique global smooth solution.
We mainly consider solutions in the class~$\Cisc(\scrM, S\scrM)$ of smooth sections
with spatially compact support. On such solutions, one has the scalar product
\beq \label{print}
(\psi_m | \phi_m)_m = 2 \pi \int_\scrN \Sl \psi_m \,|\, \nu^j \gamma_j\, \phi_m \Sr_x\: d\mu_\scrN(x) \:,
\eeq
where~$\nu$ is the future-directed normal on~$\scrN$
(due to current conservation, the scalar product is
in fact independent of the choice of~$\scrN$; for details see~\cite[Section~2]{finite}).
Forming the completion gives the Hilbert space~$(\H_m, (.|.)_m)$.

We shall also work with the Hilbert space of families of solutions of the Hilbert space
defined as follows. We consider the mass parameter in a bounded open interval, $m \in I := (m_L, m_R)$
with~$0 \not\in I$.
For a given Cauchy surface~$\scrN$, we consider a function~$\psi_\scrN(x,m) \in S_x\scrM$
with~$x \in \scrN$ and~$m \in I$. We assume that this wave function is smooth and has
compact support in both variables, $\psi_\scrN \in C^\infty_0(\scrN \times I, S\scrM)$.
For every~$m \in I$, we let~$\psi(.,m)$ be the solution of the Cauchy problem for initial data~$\psi_\scrN(.,m)$,
\beq \label{cauchy2}
(\Dir - m) \,\psi(x,m) = 0 \:,\qquad \psi(x,m) = \psi_\scrN(x,m)  \;\; \forall\: x \in \scrN \:.
\eeq
Since the solution of the Cauchy problem is smooth and depends smoothly on parameters,
we know that~$\psi \in C^\infty(\scrM \times I, S\scrM)$.
Moreover, due to finite propagation speed, $\psi(.,m)$ has spatially compact support.
Finally, the solution is clearly compactly supported in the mass parameter~$m$.
We summarize these properties by writing
\beq \label{CscmH}
\psi \in \Cisco(\scrM \times I, S\scrM) \:,
\eeq
where~$\Cisco(\scrM \times I, S\scrM)$ denotes the smooth wave functions with spatially compact support which
are also compactly supported in~$I$. 
We often denote the dependence on~$m$ by a subscript, $\psi_m(x) := \psi(x,m)$.
Then for any fixed~$m$, we can take the scalar product~\eqref{print}. 
On families of solutions~$\psi, \phi \in \Cisco(\scrM \times I, S\scrM)$ of~\eqref{cauchy2},
we introduce a scalar product by integrating over the mass parameter,
\beq \label{spm}
( \psi | \phi) := \int_I (\psi_m | \phi_m)_m \: dm
\eeq
(where~$dm$ is the Lebesgue measure). Forming the completion gives the
Hilbert space~$(\H, (.|.))$. It consists of measurable functions~$\psi(x,m)$
such that for almost all~$m \in I$, the function $\psi(.,m)$ is a weak solution of the Dirac
equation which is square integrable over any Cauchy surface.
Moreover, this spatial integral is integrable over~$m \in I$, so that
the scalar product~\eqref{spm} is well-defined. We denote the norm on~$\H$
by~$\| . \|$. The Hilbert space~$\H$ can be regarded as the {\em{direct integral}} of the
Hilbert spaces~$\H_m$, sometimes denoted alternatively by
\[ \H = L^2(I, \H_m; dm) = \int_I^\oplus \H_m\: dm \:. \]
Our procedure clarifies that the space~$\Cisco(\scrM \times I, S\scrM)$ is dense in~$\H$.

\subsection{The Dirac Equation in the Exterior Schwarzschild Geometry} \label{secschwarzschild}
In Schwarz\-schild coordinates, the line element of the Schwarzschild geometry
takes the form
\[ 
ds^2 = g_{jk}\:dx^j \,dx^k = \frac{\Delta}{r^2} \: dt^2
- \frac{r^2}{\Delta}\:dr^2 - r^2\: d \vartheta^2 - r^2\: \sin^2 \vartheta\: d\varphi^2\:, \]
where
\[ \Delta := r^2 - 2M r \:, \]
and~$M>0$ is the mass of the black hole. The zero~$r_1:=2M$ of~$\Delta$ defines
the event horizon. We here restrict attention to the {\em{exterior region}}
outside the event horizon. Thus the coordinates~$(t,r, \vartheta, \varphi)$ are in the range
\[ -\infty<t<\infty,\;\;\; r_1<r<\infty,\;\;\; 0<\vartheta<\pi,\;\;\; 0<\varphi<2\pi \:. \]
The exterior region is globally hyperbolic. The surfaces of constant coordinate time~$t$
form a foliation by Cauchy surfaces.

In~\cite{kerr, tkerr} the Dirac equation is computed in the Kerr geometry and
the solution of the Cauchy problem is expressed in terms of the radial and angular ODEs arising in the
separation of variables. In the remainder of the preliminaries, we recall
a few steps of the construction, specialized to the exterior Schwarzschild geometry.
We choose the pseudo-orthonormal frame
\[ u_0 = -\frac{r}{\sqrt{\Delta}}\: \frac{\partial}{\partial t} \:,\qquad
u_1 = \frac{1}{r}\: \frac{\partial}{\partial \vartheta} \:,\qquad
u_2 = \frac{1}{r \,\sin \vartheta}\: \frac{\partial}{\partial \varphi} \:,\qquad
u_3 = \frac{\sqrt{\Delta}}{r}\: \frac{\partial}{\partial r} \:. \]
For the Dirac operator we make the ansatz
\[ \Dir = i G^j \partial_j + B \:. \]
In order to satisfy the anti-commutation relations
\[ g^{jk}(x) \:\1_{S_x \scrM} = \frac{1}{2} \:\big\{G^j(x),\:G^k(x) \big\} \:, \]
we choose~$G^j(x) = u^j_a(x) \,\gamma^a$, where~$\gamma^a$ are the usual
Dirac matrices in the Weyl representation. More precisely, we set
\[ G^t(x) = -\frac{r}{\sqrt{\Delta}}\: \gamma^0 \:,\quad
G^\vartheta(x) = \frac{1}{r}\: \gamma^1 \:,\quad
G^\varphi(x) = \frac{1}{r \,\sin \vartheta}\: \gamma^2 \:,\quad
G^r(x) = \frac{\sqrt{\Delta}}{r}\: \gamma^3 \:, \]
where
\[ \gamma^0 = \begin{pmatrix} 0 & \1 \\ \1 & 0 \end{pmatrix} , \quad
\vec{\gamma} = \begin{pmatrix} 0 & \vec{\sigma} \\ -\vec{\sigma} & 0 \end{pmatrix} \]
(and~$\vec{\sigma}$ are the Pauli matrices).
In order to arrange that these matrices are symmetric with respect to the spin scalar
product~$\Sl .|. \Sr_x$ in~\eqref{printintro} and~\eqref{stipintro}, we choose
\beq \label{sspdef}
\Sl \psi | \phi \Sr_x := -\la \psi, \begin{pmatrix} 0 & \1 \\ \1 & 0 \end{pmatrix} \phi \ra_{\C^4}
\eeq
(here the minus sign is a good convention because then the
inner product~$\Sl . | G^t . \Sr_x$ is {\em{positive}} definite).
The corresponding zero-order term~$B$ in the Dirac operator is given by (see also the
general method for diagonal metrics in~\cite[Proposition~9.1]{topology})
\begin{align*}
B &= \frac{i}{2 \sqrt{|\det g|}}\: \partial_j \left( \sqrt{|\det g|}\, G^j \right)
= \frac{i}{2 r^2 \sin \vartheta}\: \partial_j \left( r^2 \sin \vartheta\, G^j \right) \\
&= \frac{i}{2 \sin \vartheta}\: \partial_\vartheta \big( \sin \vartheta\, G^\vartheta \big)
+ \frac{i}{2 r^2}\: \partial_r \big( r^2\, G^r \big)
= \frac{i \cot \vartheta}{2 r}\:\gamma^1 + \frac{i\partial_r \big( r\, \sqrt{\Delta} \big)}{2 r^2}\: \, \gamma^3 \:.
\end{align*}
The resulting Dirac operator takes the form
\begin{align*}
\Dir &= \begin{pmatrix} 0 & 0 & \alpha_+ & \beta_+ \\
0 & 0 & \beta_- & \alpha_- \\
\alpha_- & -\beta_+ & 0 & 0 \\
-\beta_- & \alpha_+ & 0 & 0 \end{pmatrix} \qquad \text{with} \\
\beta_\pm &= \frac{i}{r} \left( \frac{\partial}{\partial
\vartheta} + \frac{\cot \vartheta}{2} \right) \pm
\frac{1}{r \sin \vartheta} \:\frac{\partial}{\partial \varphi} \\
\alpha_\pm &= -\frac{ir} {\sqrt{\Delta}} \:\frac{\partial}{\partial t}
\pm \frac{\sqrt{\Delta}}{r} \left( i \frac{\partial}{\partial r}
\:+\: i \:\frac{r-M}{2 \Delta} \:+\: \frac{i}{2r} \right) .
\end{align*}

\subsection{Separation of the Dirac Equation} \label{secsep}
In preparation, we let $S(r)$ and
$\Gamma(r)$ be the diagonal matrices
\[  S = \Delta^{\frac{1}{4}} \:\sqrt{r}\: \1_{\C^4} \:, \qquad
\Gamma = -i r \:{\mbox{diag}} \left( 1,\: -1,\: -1,\: 1 \right) \:. \]
Then the transformed wave function
\beq \label{psitrans}
\Psi = S \,\psi
\eeq
satisfies the Dirac equation
\beq \label{eq:11}
\Gamma S\:(\Dir-m)\:S^{-1} \:\Psi = 0 \:.
\eeq
Moreover,
\[ \Gamma S\:(\Dir-m)\:S^{-1} = {\mathcal{R}} + {\mathcal{A}} \]
with
\begin{align*}
{\mathcal{R}} &= \begin{pmatrix} imr & 0 & \sqrt{\Delta}\: {\mathcal{D}}_+ & 0 \\
0 & -imr & 0 & \sqrt{\Delta} \:{\mathcal{D}}_- \\
\sqrt{\Delta} \:{\mathcal{D}}_- & 0 & -imr & 0 \\
0 & \sqrt{\Delta}\:{\mathcal{D}}_+ & 0 & imr \end{pmatrix} \\
{\mathcal{A}} &= \begin{pmatrix} 0 & 0 & 0 & {\mathcal{L}}_+ \\
0 & 0 & -{\mathcal{L}}_- & 0 \\
0 & {\mathcal{L}}_+ & 0 & 0 \\
-{\mathcal{L}}_- & 0 & 0 & 0 \end{pmatrix}
\end{align*}
and
\[ {\mathcal{D}}_\pm = \frac{\partial}{\partial r} \mp
\frac{r^2}{\Delta}\:\frac{\partial}{\partial t} \:,\qquad
{\mathcal{L}}_\pm = \frac{\partial}{\partial \vartheta} + \frac{\cot
\vartheta}{2} \mp \frac{i}{\sin \vartheta}
\:\frac{\partial}{\partial \varphi} \:. \]

For the separation of the Dirac equation, we first employ for~$\Psi$ the ansatz
\beq
\Psi(t,r,\vartheta,\varphi) = e^{-i \omega t} \:e^{-i k \varphi} \:
\Phi(r, \vartheta) \:,\qquad \text{with~$\omega \in \R, \:k \in \Z+\frac{1}{2}$} \:. \label{eq:16a}
\eeq
Next, for the function~$\Phi$ we make the ansatz
\beq
\Phi(r,\vartheta) =
\left( \begin{array}{c} X_-(r) \:Y_-(\vartheta) \\
X_+(r) \:Y_+(\vartheta) \\
X_+(r) \:Y_-(\vartheta) \\
X_-(r) \:Y_+(\vartheta) \end{array} \right), \label{eq:16b}
\eeq
composed of radial functions~$X_\pm(r)$ and angular functions~$Y_\pm(\vartheta)$.
By substituting (\ref{eq:16a}) and (\ref{eq:16b}) into the transformed
Dirac equation (\ref{eq:11}), we obtain the eigenvalue problems
\[ {\mathcal{R}} \:\Psi = \lambda \:\Psi \:,\qquad
 {\mathcal{A}} \:\Psi = -\lambda \:\Psi \:, 
\]
under which the Dirac equation (\ref{eq:11}) decouples into the system of ODEs
\begin{align}
\left( \begin{array}{cc} \sqrt{\Delta} \:{\mathcal{D}}_+ & imr - \lambda \\
-imr - \lambda & \sqrt{\Delta} \:{\mathcal{D}}_- \end{array} \right)
\left( \begin{array}{c} X_+ \\ X_- \end{array} \right) &= 0
\label{eq:21a} \\
\left( \begin{array}{cc} {\mathcal{L}}_+ & \lambda \\
\lambda & -{\mathcal{L}}_- \end{array} \right)
\left( \begin{array}{c} Y_+ \\ Y_- \end{array} \right) &= 0
\label{eq:21b}
\end{align}
with
\beq \label{DLode}
{\mathcal{D}}_\pm = \frac{\partial}{\partial r} \pm i \omega \,\frac{r^2}{\Delta}\:,\qquad
{\mathcal{L}}_\pm = \frac{\partial}{\partial \vartheta} + \frac{\cot
\vartheta}{2} \mp \frac{k}{\sin \vartheta} \:.
\eeq

\subsection{Fundamental Solutions and their Asymptotics} \label{secfs}
The angular equation~\eqref{eq:21b} with~${\mathcal{L}}_\pm$ according to~\eqref{DLode}
does not involve~$\omega$. For any~$k \in \Z+\frac{1}{2}$, it can be regarded as an eigenvalue equation
for the angular function~$Y$. 
This eigenvalue equation can be solved by the so-called spin-weighted
spherical harmonics (for details see~\cite{goldberg}).
We thus obtain an orthonormal eigenvector basis~$Y_{kn}$ with~$n \in \Z$
in the Hilbert space~$L^2((-1,1), d\cos \vartheta)^2$, i.e.\
\beq \label{angortho}
\big\la e^{-i k \varphi}\: Y_{kn}(\vartheta), e^{-i k' \varphi}\: Y_{k'n'}(\vartheta) \big\ra_{L^2(S^2)^2} = 
\delta_{k,k'}\: \delta_{n,n'}\:.
\eeq
We denote the corresponding eigenvalues by~$\lambda_{kn}$.

Choosing the separation constant~$\lambda=\lambda_{kn}$ as one of these eigenvalues, the
radial ODE~\eqref{eq:21a} can be written as
\beq \label{radial}
\left[ \frac{d}{du} + i \omega
\left( \begin{array}{cc} 1 & 0 \\ 0 & -1 \end{array} \right) \right] X
= \frac{\sqrt{\Delta}}{r^2} \:\left( \begin{array}{cc} 0 & \lambda-imr \\
\lambda + imr & 0 \end{array} \right) X \:,
\eeq
where for convenience we transformed
the radial variable to the Regge-Wheeler coordinate~$u \in \R$ defined by
\[  \frac{du}{dr} = \frac{r^2}{\Delta} \:. 
\]
The limit~$u \rightarrow -\infty$ describes the event horizon, whereas in the limit~$u \rightarrow \infty$
one reaches spatial infinity.
The asymptotics of the solutions of the radial ODE
have been worked out in~\cite[Lemmas~3.1 and~3.5]{tkerr}:
\begin{Lemma}
\label{lemma31}
Every solution $X$ of~\eqref{radial} 
is asymptotically as $u \rightarrow -\infty$ of the form
\[ X(u) = \begin{pmatrix} e^{-i \omega u} \:f_0^+ \\
e^{i \omega u} \:f_0^- \end{pmatrix} + R_0(u) 
\]
with~$f^\pm_0 \in \C$ and an exponentially decaying error term, i.e.\
\[ |R_0| \leq c \:e^{du} \]
with a constant~$c,d>0$ which can be chosen locally uniformly in~$\omega$ and~$|f_0^\pm|$.
\end{Lemma}

\begin{Lemma} \label{lemma3}
In the case~$|\omega|<m$, the ODE~\eqref{radial} has one fundamental solution which
decays exponentially as~$u \rightarrow \infty$, and one fundamental solution which increases
exponentially in this limit.

In the case~$|\omega|>m$, on the other hand, every solution $X$ of~\eqref{radial} 
is asymptotically as $u \rightarrow \infty$ of the form
\begin{equation}
X(u) = A \begin{pmatrix} e^{-i \Phi^\omega_m(u)} \:f_\infty^+ \\
e^{i \Phi^\omega_m(u)} \:f_\infty^- \end{pmatrix} + R_\infty(u)
    \label{eq:3s}
\end{equation}
with~$f_\infty^\pm \in \C$ and
\begin{align}
\Phi^\omega_m &= \epsilon(\omega) \left(
\sqrt{\omega^2-m^2} \:u \:+\: \frac{M m^2}
{\sqrt{\omega^2-m^2}} \:\log u \right) \label{eq:3u} \\
A &= \left( \begin{array}{cc} \cosh \Theta & \sinh \Theta \\
\sinh \Theta & \cosh \Theta \end{array} \right) \;,\qquad
\Theta = \frac{1}{4}\: \log \left| \frac{\omega-m}{\omega+m} \right|
\label{eq:3v} \\
|R_\infty| &\leq \frac{C}{u} \:, \label{eq:3w}
\end{align}
where the constant $C>0$ can be chosen locally uniformly in~$\omega$ and~$f_\infty^\pm$.
\end{Lemma}

Based on these results, we choose fundamental solutions of the Dirac equation
\beq \label{Psifund}
\Psi^{k\omega n}_{m,a} \qquad \text{with~$k \in \Z+\frac{1}{2}$, $\omega \in \R$, $n \in \Z$
and~$a=1,2$}
\eeq
as follows. We always use the separation ansatz~\eqref{eq:16a} and~\eqref{eq:16b}
with~$Y$ chosen as the angular eigenfunction~$Y_{kn}$
and the radial eigenfunction~$X$ as a solution of the radial ODE~\eqref{radial}
with~$\lambda=\lambda_{kn}$. In the case~$|\omega|< m$,
the solution for~$a=1$ decays exponentially at infinity, whereas the solution for~$a=2$
increases exponentially. We normalize the solutions such that they have norm one
at the event horizon
\[ \|f^{k \omega n}_{0,\:m,a}\|_{\C^2} = 1 \qquad \text{for~$a=1,2$} \]
(where~$f^{k \omega n}_{0,\:a} \in \C^2$ is the vector with
components denoted by~$\pm$).
In the case~$|\omega|>m$, on the other hand, we choose
the fundamental solutions with the asymptotics near the event horizon
\beq \label{fevent}
f^{k \omega n}_{0,\: m,1} = \begin{pmatrix} 1 \\ 0 \end{pmatrix} \:,\qquad
f^{k \omega n}_{0,\: m,2} = \begin{pmatrix} 0 \\ 1 \end{pmatrix} \:.
\eeq
The asymptotics of these solutions at spatial infinity is described as in~\eqref{eq:3s}
by coefficients~$f^\pm_\infty$. For clarity, we combine these so-called
{\em{transmission coefficients}} to a vector denoted by~$f^{k \omega n}_{\infty,\: m,a} \in \C^2$.

\subsection{An Integral Representation of the Dirac Propagator}
The dynamics of Dirac waves in a black hole geometry can be analyzed with different methods.
Scattering theory as developed in~\cite{nicolas95, melnyk, daude} gives information on the
behavior of the waves near the event horizon and near infinity. Moreover, it connects the asymptotic
waves by scattering operators.
Another approach are integral representations for the Dirac propagator which were derived in~\cite{kerr, tkerr}
and used to study decay rates in~\cite{decay}.
The advantage of the integral representations is that they give finer information on the
solutions of the Cauchy problem. However, the integral representations apply only in special geometries
where the Dirac equation is separable into ordinary differential equations (ODEs), namely in the
Kerr-Newman black hole geometry and special cases thereof like the Kerr, Reissner-Nordstr\"om
and Schwarzschild black holes. Here we shall work with the integral representations because this
makes it possible to work out the mass decomposition explicitly in terms of the fundamental solutions
of the radial ODE and the corresponding transmission coefficients.

More precisely, an integral representation of the Dirac propagator was derived in~\cite[Theorem~3.6]{tkerr}.
Here we write this representation as follows.
First, it was shown that every smooth
and spatially compact solution of the Dirac equation~$\psi_m \in \Cisc(\scrM, S\scrM)$ can be
represented as
\beq \label{superposition}
\psi_m(t,r,\vartheta, \varphi) = \sum_{k,n}
\int_\R d\omega \:e^{-i \omega t} \,\sum_{a=1}^2
\hat{\psi}_{m,a}^{kn}(\omega)\: \Psi^{k \omega n}_{m,a}(r,\vartheta,\varphi)
\eeq
with complex-valued functions~$\hat{\psi}_{m,a}^{kn}(\omega)$,
where we sum over~$k \in \Z +\frac{1}{2}$ and~$n \in \Z$. Here~$\Psi^{k \omega n}_{m,a}$ are the
fundamental solutions of the coupled ODEs introduced in~\eqref{Psifund}.
Moreover, in the case~$|\omega|<m$ only those fundamental solutions appear
which decay at spatial infinity. We implement this fact by always choosing
\beq \label{decayfund}
\hat{\psi}_{m,2}^{kn}(\omega) = 0 \qquad \text{for all~$\omega \in (-m,m)$ and all~$k,n$}\:.
\eeq
Second, the integral representation in~\cite[Theorem~3.6]{tkerr} also gives
explicit formulas for the functions~$\hat{\psi}_{m,a}^{kn}$ in~\eqref{superposition}
in terms of the initial data~$\Psi_0 \in C^\infty_0(\scrN, S\scrM)$
(where~$\scrN$ is the hypersurface~$\{t=0\}$). Indeed,
\beq \label{hatpsi}
\hat{\psi}_{m,a}^{kn}(\omega) = \frac{1}{2 \pi^2} \sum_{b=1}^2
t^{k \omega n}_{ab} \:(\Psi^{k \omega n}_{m,b} \,|\, \psi_m )_m \big|_{t=0}\:,
\eeq
where~$(.|.)_m$ is again the scalar product~\eqref{print}, and the coefficients~$t^{k \omega n}_{ab}$
can be expressed explicitly in terms of the transmission coefficients
(for the prefactor $1/(2 \pi^2)$ one must keep in mind that the scalar product~\eqref{print}
involves a factor~$2 \pi$).

\section{Derivation of the Mass Decomposition} \label{secmassdecomp}
For the analysis of mass oscillations, we consider
a variable mass parameter~$m$ in an interval~$I=(m_L, m_L)$ with~$0 \not\in I$.
Following the general procedure in~\cite{infinite}, we want to choose the {\em{domain}}~$\H^\infty$
as a dense subspace of~$\H$ spanned by families of Dirac solutions which are ``nice'' in the sense
that it is easy to handle technical issues like verifying differentiability, integrability and
interchangeability of limits and integrals. Here it is most convenient to make the following choice:

\begin{Def} \label{defHinf}
The {\em{domain}}~$\H^\infty \subset \H$
is chosen as the space of all Dirac solutions
of the form~\eqref{superposition} and~\eqref{decayfund} which satisfy the following conditions:
\bitem
\item[\rm{(i)}] The functions~$\hat{\psi}^{kn}_{m,a}(\omega)$ in~\eqref{superposition}
vanish identically for almost all~$k \in \Z+\frac{1}{2}$ and~$n \in \Z$.
\item[\rm{(ii)}] For all~$k \in \Z+\frac{1}{2}$, $n \in \Z$ and~$a \in \{1,2\}$,
the functions~$\hat{\psi}^{kn}_{m,a}(\omega)$ are smooth and compactly supported in~$\omega$ and~$m$.
Moreover, they are supported away from~$\omega=\pm m$, i.e.\
\beq \label{supp}
\supp \hat{\psi}^{kn}_{.,a}(.) \;\subset \; \big\{ (\omega, m) \in \R \times I \quad\text{with}\quad \omega \neq \pm m \big\} \:.
\eeq
\eitem
\end{Def} \noindent
We clarify this definition with two remarks.
We first note that, by approximation, the integral representation~\eqref{superposition}
can be extended to wave functions which do not have spatially compact support.
In particular, choosing~$\hat{\psi}^{kn}_{m,a}(\omega)$ according to Definition~\ref{defHinf}~(i) and~(ii),
for all~$m \in I$ one gets a well-defined Dirac solution in~$\H_m$.
The second remark concerns the condition~\eqref{supp}.
This condition states that the waves in~$\H^\infty$ vanish identically for frequencies in
a neighborhood of~$\omega = \pm m$. In other words, we disregard Dirac solutions which asymptotically
at infinity have zero momentum. This assumption is a major technical simplification.
It is justified by the fact that~$\H^\infty$ is dense in~$\H$, as can be verified
in various ways: One method is to work again with the integral representation~\eqref{superposition}
and to approximate the functions~$\hat{\psi}^{kn}_{m,a}$ coming from a
solution~$\psi \in \Cisco(\scrM, S\scrM)$ by functions which vanish in a neighborhood of~$\omega=\pm m$.
More abstractly, the denseness of~$\H^\infty$ follows from the fact that the Dirac Hamiltonian has no
point spectrum. Using methods of scattering theory, this follows
from the asymptotic completeness results in~\cite{nicolas95, melnyk}.

\subsection{Integral Representation of the Scalar Product}
After performing the trans\-for\-ma\-tion~\eqref{psitrans}, the scalar product~\eqref{print}
with the spin scalar product according to~\eqref{sspdef} takes for all~$\psi, \phi \in \H^\infty$ the form
\[ (\psi_m | \phi_m)_m = 2 \pi \int_{r_1}^\infty dr\int_{-1}^1 d\cos \vartheta
\int_0^{2 \pi} d\varphi \; \frac{r^2}{\Delta} \; \big\la \Psi_m, \Phi_m \big\ra_{\C^4} \:. \]
Employing~\eqref{superposition} as well as the separation ansatz~\eqref{eq:16b},
one can use the orthogonality of the angular eigenfunctions~\eqref{angortho} to obtain
\begin{align*}
(\psi_m | \phi_m)_m &= 4 \pi \sum_{k,n} \int_{r_1}^\infty dr\: \frac{r^2}{\Delta} 
\int_{-\infty}^\infty d\omega \int_{-\infty}^\infty d\omega' \:e^{i (\omega-\omega') t} \\
&\qquad\qquad \times \sum_{a, a'=1}^2 \overline{\hat{\psi}_{m,a}^{kn}(\omega)}\: \hat{\phi}_{m,a'}^{kn}(\omega')\: 
\Big\la X^{k \omega n}_{m,a}(r), X^{k \omega' n}_{m,a'}(r) \Big\ra_{\C^2} \:,
\end{align*}
where the functions~$\hat{\phi}^{kn}_{m,a}, \hat{\psi}^{kn}_{m,a}$ are defined by~\eqref{hatpsi}
(note that a factor of two arises because a four-spinor involves~$X$ twice).
This representation has the disadvantage that, due to the factor~$e^{i (\omega-\omega') t}$,
current conservation is not apparent. Therefore, the representations derived
in the following lemma are more useful (these representations were first used in~\cite[Section~9]{decay}).
\begin{Lemma} \label{lemmaprintrep}
For any Dirac solutions~$\psi_m, \phi_m \in \Cisc(\scrM, S\scrM)$,
their scalar product can be written in the alternative forms
\begin{align}
(\psi_m | \phi_m)_m &= 2 \pi^2 \sum_{k,n}
\int_\R \sum_{a,b=1}^2 \big(T^{kn}(\omega)^{-1}\big)^{ab}\;
\overline{\hat{\psi}^{kn}_{m,a}(\omega)} \:\hat{\phi}^{kn}_{m,b}(\omega)\: d\omega \label{rep1} \\
&= \frac{1}{2 \pi^2} 
\sum_{k,n} \int_\R \sum_{a,b=1}^2
t^{k \omega n}_{ab} \:(\psi_m \,|\, \Psi^{k \omega n}_{m,a})_m \:(\Psi^{k \omega n}_{m,b} \,|\, \phi_m)_m \big|_{t=0}\:
d\omega \:,
\label{rep2}
\end{align}
where~$T^{kn}(\omega)$ is the $2 \times 2$-matrix with entries~$(t^{k \omega n}_{ab})_{a,b=1,2}$.
\end{Lemma}
\Proof Combining~\eqref{superposition} and~\eqref{hatpsi} and setting~$t=0$, we obtain
\[ \psi_m|_{t=0} (r,\vartheta,\varphi) = \frac{1}{2 \pi^2} \sum_{k,n} \int_\R \sum_{a,b=1}^2
t^{k \omega n}_{ab} \:(\Psi^{k \omega n}_{m,b} \,|\, \psi_m)_m \big|_{t=0}\: \Psi^{k \omega n}_{m,a}(r,\vartheta,\varphi) 
\:d\omega \:. \]
Taking the scalar product with another Dirac solution~$\phi_m$ gives~\eqref{rep2}.
Again applying~\eqref{hatpsi} gives~\eqref{rep1}.
\QED

\subsection{Mass Decomposition of the Space-Time Inner Product}
After the transformation~\eqref{psitrans}, the space-time inner product~\eqref{stip} 
(with the spin scalar product according to~\eqref{sspdef}) becomes
\[ \bra \psi|\phi \ket 
= -\int_{-\infty}^\infty dt \int_{r_1}^\infty dr\int_{-1}^1 d\cos \vartheta
\int_0^{2 \pi} d\varphi \; \frac{r}{\sqrt{\Delta}} \; \Big\la \Psi,
\begin{pmatrix} 0 & \1 \\ \1 & 0 \end{pmatrix} \Phi \Big\ra_{\C^4} \:. \]
Applying~\eqref{superposition} as well as the separation ansatz~\eqref{eq:16b},
one can again use the orthogonality of the angular eigenfunctions~\eqref{angortho} 
as well as Plancherel's theorem to obtain
for all~$\psi, \phi \in \H^\infty$
\begin{align}
\bra \p \psi \,|\, \p \phi \ket &= -2 \sum_{k,n} \int_{-\infty}^\infty dt \int_{r_1}^\infty dr\: \frac{r}{\sqrt{\Delta}} 
\int_I dm \int_I dm' 
\int_{-\infty}^\infty d\omega \int_{-\infty}^\infty d\omega' \:e^{i (\omega-\omega') t} \notag \\
&\qquad \times \sum_{a, a'=1}^2 
\overline{\hat{\psi}_{m,a}^{kn}(\omega)}\: \hat{\phi}_{m',a'}^{kn}(\omega')\: 
\Big\la X^{k \omega n}_{m,a}(r), \begin{pmatrix} 0 & 1 \\ 1 & 0 \end{pmatrix}
X^{k \omega' n}_{m',a'}(r) \Big\ra_{\C^2} \notag \\
&= -4 \pi \sum_{k,n} \int_{r_1}^\infty dr\: \frac{r}{\sqrt{\Delta}} 
\int_I dm \int_I dm' \int_{-\infty}^\infty d\omega \notag \\
&\qquad \times \sum_{a, a'=1}^2 
\overline{\hat{\psi}_{m,a}^{kn}(\omega)}\: \hat{\phi}_{m',a'}^{kn}(\omega)\: 
\Big\la X^{k \omega n}_{m,a}(r), \begin{pmatrix} 0 & 1 \\ 1 & 0 \end{pmatrix}
X^{k \omega n}_{m',a'}(r) \Big\ra_{\C^2} \label{stipsep}
\end{align}
(note that again a factor of two arises because a four-spinor involves~$X$ twice).
Since in this formula as well as in the formulas of Lemma~\ref{lemmaprintrep}
we get a pairing only between wave functions with the same angular momentum,
in what follows we may restrict attention to a single angular momentum mode.
Consequently, from now on we always leave out the sums over~$k$ and~$n$ 
and omit the indices~$k$ and~$n$.

The $r$-integration in~\eqref{stipsep} can be carried out, giving the following result.
\begin{Thm} \label{thmstiprep} {\bf{(mass decomposition of the space-time inner product)}} \\
Restricting attention to one angular momentum mode, for all~$\psi, \phi \in \H^\infty$
the following identity holds,
\begin{align*}
\bra \p \psi \,|\, \p \phi \ket &= 4 \pi^2 \int_I dm
\int_{\R \setminus [-m,m]} \!\!\!\!\!\!\!\!\!\epsilon(\omega)\: d\omega \sum_{a, a'=1}^2
\overline{\hat{\psi}_{m,a}(\omega)}\: \hat{\phi}_{m,a}(\omega) \:
\la f_{\infty,\: m,a}^{\omega}, \,f_{\infty,\:m, a'}^{\omega} \ra_{\C^2} \\
&\quad\,-4 \pi i \int_I dm \int_I dm' \: \frac{\text{\rm{PP}}}{m-m'} \int_{-\infty}^\infty d\omega \\
&\qquad\qquad\qquad \times \sum_{a,a'=1}^2
\overline{\hat{\psi}_{m,a}(\omega)}\: \hat{\phi}_{m',a'}(\omega) \;
\la f^{\omega}_{0,\: m,a} , \begin{pmatrix} 1 & 0 \\ 0 & -1 \end{pmatrix} f^{\omega}_{0,\: m',a'} \ra_{\C^2} \:,
\end{align*}
where~$\hat{\psi}_{m,a}$ and~$\hat{\phi}_{m',a'}$ are the functions in the representation~\eqref{superposition}
(which always satisfy the condition~\eqref{decayfund}). Moreover, 
$f^{\omega}_{0,\: m,a}$ and~$f^{\omega}_{\infty,\: m,a}$
describe the asymptotics of the radial fundamental solutions near the
event horizon and at infinity (see Lemmas~\ref{lemma31} and~\ref{lemma3}).
\end{Thm}

The remainder of this section is devoted to the proof of this theorem.
Its physical significance will be explained afterwards in Section~\ref{secinterpret}.
We begin with preparatory lemmas.
\begin{Lemma} \label{lemmapseudo}
In the range~$|\omega|>m$, the transmission coefficients~$f^\omega_{\infty,\:m,a}$ satisfy
the relations
\begin{align}
\big|f^{\omega+}_{\infty,\:m,a}\big|^2 - \big|f^{\omega-}_{\infty,\:m,a}\big|^2 &= 
\left\{ \begin{array}{cl} 1 & \text{if~$a=1$} \\
-1 & \text{if~$a=2$} \:. \end{array} \right. \label{fpseudo} \\
\big\|f^\omega_{\infty,\:m,1} \big\|_{\C^2} &= \big\|f^\omega_{\infty,\:m,2} \big\|_{\C^2} 
\geq 1 \:. \label{feuclid}
\end{align}
\end{Lemma}
\Proof The form of the matrices in the ODE~\eqref{radial} imply that 
(see~\cite[proof of Lemma~3.3]{tkerr}),
\beq \label{Xnormdiff}
\frac{d}{du} \left(|X_+|^2 - |X_-|^2 \right) = 0 \:.
\eeq
Hence, in view of the asymptotics near the event horizon~\eqref{fevent}, it follows that
\[ |X_+(u)|^2 - |X_-(u)|^2 = \left\{ \begin{array}{cl} 1 & \text{if~$a=1$} \\
-1 & \text{if~$a=2$} \end{array} \right. \qquad \text{for all~$u \in \R$}\:. \]
Using the asymptotics at infinity of Lemma~\ref{lemma3}, we conclude that
\[ \la A f^{\omega}_{\infty,\: m,a} , \begin{pmatrix} 1 & 0 \\ 0 & -1 \end{pmatrix} A f^{\omega}_{\infty,\: m,a} \ra_{\C^2} 
= \left\{ \begin{array}{cl} 1 & \text{if~$a=1$} \\
-1 & \text{if~$a=2$} \:. \end{array} \right. \]
It is useful to regard the bilinear form on the left as an indefinite inner product
generated by the matrix~$\text{diag}(1,-1)$. By direct computation one verifies
that the matrix~$A$ in~\eqref{eq:3v} is unitary with respect to this inner product.
This gives~\eqref{fpseudo}.

In order to derive~\eqref{feuclid}, we make use of the fact that the identity~\eqref{Xnormdiff}
holds for any linear combination of our fundamental solutions.
This implies that the transmission coefficients are pseudo-orthonormal with respect to
the indefinite inner product generated by the matrix~$\text{diag}(1,-1)$, i.e.
\beq \label{pseudoortho}
\la f^{\omega}_{\infty,\: m,a} , \begin{pmatrix} 1 & 0 \\ 0 & -1 \end{pmatrix} f^{\omega}_{\infty,\: m,b} \ra_{\C^2} 
= \delta_{ab} \:\times \left\{ \begin{array}{cl} 1 & \text{if~$a=1$} \\
-1 & \text{if~$a=2$} \:. \end{array} \right.
\eeq
As a consequence, the transmission coefficients can be parametrized as
\beq \label{fparam}
f^{\omega}_{\infty,\: m,1} = \begin{pmatrix} e^{i \beta}\, \cosh \vartheta \\ 
e^{i \gamma}\, \sinh \vartheta \end{pmatrix} \:, \qquad
f^{\omega}_{\infty,\: m,2} = e^{i \delta} \begin{pmatrix} e^{i \beta}\, \sinh \vartheta \\
e^{i \gamma}\, \cosh \vartheta \end{pmatrix} \:,
\eeq
with four real parameters~$\vartheta, \beta, \gamma, \delta$.
Indeed, the left equation is a general parametrization of a vector satisfying the normalization~\eqref{fpseudo}.
The right equation, on the other hand, parametrizes a general vector which again satisfies the normalization~\eqref{fpseudo} and is orthonormal to the first vector
with respect to the inner product in~\eqref{pseudoortho}.

In the parametrization~\eqref{fparam}, the relations in~\eqref{feuclid} are verified immediately by a short computation.
\QED

\begin{Lemma} \label{lemmadecayinf} For all~$\psi \in \H^\infty$ and all~$\omega \in \R$,
\[ \int_I dm \sum_{a=1}^2 
\hat{\psi}_{m,a}(\omega)\: X^{\omega}_{m,a}(r) = \O \Big( \frac{1}{r} \Big) \:. \]
\end{Lemma}
\Proof We first point out that, due to the support assumption~\eqref{supp},
the function~$\hat{\psi}_{m,a}(\omega)$ vanishes in a neighborhood of~$m = \pm \omega$.
This has the technical advantage that the transformation functions in Lemma~\ref{lemma3}
are smooth in~$m$. Moreover, we may treat the cases~$|\omega|<m$ and~$|\omega|>m$ separately.

In the case~$|\omega|<m$, the fundamental solution~$X^\omega_{m,1}$ decays
exponentially, giving the desired decay for large~$r$.
In the remaining case~$|\omega|>m$, we may clearly disregard the error term~$R_\infty$ in~\eqref{eq:3s}.
The remaining summand in~\eqref{eq:3s}
is smooth and involves the oscillatory factors~$e^{\pm i \Phi^\omega_m(u)}$. Using that
the Fourier transform of a smooth function has rapid decay, the resulting term
decays even rapidly in~$r$.
\QED

In the next lemma, the Dirac operator is ``integrated by parts'' in the space-time inner product.
As in~\cite[Section~3.1]{infinite}, we denote the operator of multiplication with the mass parameter by~$T$,
\[ T \::\: \H^\infty \rightarrow \H^\infty \:,\qquad (T \psi)_m = m \,\psi_m \:. \]

\begin{Lemma} \label{lemmapint}
For all~$\psi, \phi \in \H^\infty$,
\beq \label{pTint}
\begin{split}
&\bra \p \psi \,|\, \p T \phi \ket - \bra \p T \psi \,|\, \p \phi \ket = 4 \pi i \int_I dm \int_I dm' \\
&\qquad \times
\int_{-\infty}^\infty \sum_{a,a'=1}^2 \overline{\hat{\psi}_{m,a}(\omega)}\: \hat{\phi}_{m',a'}(\omega) \;
\la f^{\omega}_{0,\: m,a} , \begin{pmatrix} 1 & 0 \\ 0 & -1 \end{pmatrix} f^{\omega}_{0,\: m',a'} \ra_{\C^2}\:
d\omega \:.
\end{split}
\eeq
\end{Lemma}
\Proof We again restrict attention to a single angular momentum mode.
We write the radial equation~\eqref{radial} as
\[ \D X = m X \]
with the ``radial Dirac operator''
\[ \D = \frac{i \sqrt{\Delta}}{r}\: \begin{pmatrix} 0 & -1 \\ 1 & 0 \end{pmatrix}\: \frac{d}{dr}
+ \frac{\lambda}{r}\: \begin{pmatrix} i & 0 \\ 0 & -i \end{pmatrix} 
- \frac{\omega r}{\sqrt{\Delta}}\: \begin{pmatrix} 0 & 1 \\ 1 & 0 \end{pmatrix} \:. \]
Since all the matrices in this equation are symmetric with respect to the ``separated spin scalar product''
\[ \Sl .|. \Sr := -\Big\la .\,, \begin{pmatrix} 0 & 1 \\ 1 & 0 \end{pmatrix} . \Big\ra_{\C^2} \:, \]
the radial Dirac operator is symmetric with respect to the inner product~$\bra .|. \ket$.
This means that on the left side of~\eqref{pTint} our task is to compute the boundary terms
at infinity and on the horizon, i.e.\ using~\eqref{stipsep}
\begin{align*}
&\bra \p \psi \,|\, \p T \phi \ket - \bra \p T \psi \,|\, \p \phi \ket 
= -4 \pi i \:\bigg( \int_I dm \int_I dm' \int_{-\infty}^\infty d\omega \\
&\qquad \times
\sum_{a, a'=1}^2 
\overline{\hat{\psi}_{m,a}(\omega)}\: \hat{\phi}_{m',a'}(\omega)\: 
\Big\la X^{\omega}_{m,a}(r), \begin{pmatrix} 1 & 0 \\ 0 & -1 \end{pmatrix}
X^{\omega'}_{m',a'}(r) \Big\ra_{\C^2} \bigg) \bigg|^{r=\infty}_{r=r_1}\:. 
\end{align*}
The boundary terms at spatial infinity vanish in view of Lemma~\ref{lemmadecayinf}.
On the event horizon, on the other hand, we can use the
asymptotics in Lemma~\ref{lemma31}, giving the result.
\QED

We next compute the singular contribution to the space-time inner product at~$m=m'$.
\begin{Lemma} \label{lemmasing}
For all~$\psi, \phi \in \H^\infty$,
\begin{align}
&\!\!\bra \p \psi \,|\, \p \phi \ket \notag \\
&\!\!\!= 4 \pi^2 \int_I dm\:
\int_{\R \setminus [-m,m]} \epsilon(\omega)\: d\omega \sum_{a, a'=1}^2
\overline{\hat{\psi}_{m,a}(\omega)}\: \hat{\phi}_{m,a'}(\omega) \:
\la f_{\infty,\: m,a}^{\omega}, \,f_{\infty,\:m, a'}^{\omega} \ra_{\C^2} \label{print1} \\
&\;\; -4 \pi i \int_I dm \int_I dm' \: \frac{\text{\rm{PP}}}{m-m'} \notag \\
&\qquad\qquad\qquad\qquad \times \int_{\R \setminus [-m,m]}
\Big( \overline{\hat{\psi}_{m,1}(\omega)}\: \hat{\phi}_{m',1}(\omega)
- \overline{\hat{\psi}_{m,2}(\omega)}\: \hat{\phi}_{m',2}(\omega) \Big) \:d\omega \label{print22} \\
&\;\; +\int_I dm \int_I dm' \: h(m,m') \label{print3}
\end{align}
with a bounded function~$h \in L^\infty(I \times I)$.
\end{Lemma}
\Proof Considering again one angular momentum mode, \eqref{stipsep} becomes
\begin{align}
\bra \p \psi \,|\, \p \phi \ket &= -4 \pi \int_{r_1}^\infty dr\: \frac{r}{\sqrt{\Delta}} 
\int_I dm \int_I dm' \int_{-\infty}^\infty d\omega \notag \\
&\qquad\quad \times
\sum_{a, a'=1}^2 \overline{\hat{\psi}_{m,a}(\omega)}\: \hat{\phi}_{m',a}(\omega)\: 
\Big\la X^{\omega}_{m,a}(r), \begin{pmatrix} 0 & 1 \\ 1 & 0 \end{pmatrix}
X^{\omega}_{m',a'}(r) \Big\ra_{\C^2} \:. \label{forder}
\end{align}
According to Lemma~\ref{lemma31} and the fact that the weight~$r/\sqrt{\Delta}$ is integrable,
the $r$-integral converges near the event horizon for any fixed~$\omega$, $m$ and~$m'$,
uniformly in these parameters. 
Near spatial infinity, on the other hand, the plane-wave asymptotics of Lemma~\ref{lemma3} shows that
the $r$-integral in general does {\em{not}} converge for fixed~$\omega$, $m$ and~$m'$,
but the~$r$-integration is well-defined only if the integrals are performed in the order given in~\eqref{forder}.
However, we want to analyze the $r$-integral first. To this end, we insert a convergence-generating
factor~$e^{-\varepsilon r}$, making it possible to interchange the integrals
(in the end, we shall take the limit~$\varepsilon \searrow 0$ in the distributional sense).
We thus obtain
\begin{align}
\bra \p \psi \,|\, \p \phi \ket = -4 \pi \lim_{\varepsilon \searrow 0} &
\int_I dm \int_I dm' \int_{-\infty}^\infty d\omega \sum_{a, a'=1}^2 \overline{\hat{\psi}_{m,a}(\omega)}\:
\hat{\phi}_{m',a'}(\omega) \notag \\
&\times \int_{r_1}^\infty \frac{r}{\sqrt{\Delta}} \:
\Big\la X^{\omega}_{m,a}(r), \begin{pmatrix} 0 & 1 \\ 1 & 0 \end{pmatrix}
X^{\omega}_{m',a'}(r) \Big\ra_{\C^2} \: e^{-\varepsilon r} \:dr\:. \label{ppex}
\end{align}
In the case~$|\omega|<m$, the fundamental solution decays exponentially at infinity,
so that the resulting contribution can be absorbed into the function~$h(m,m')$.
In the remaining case~$|\omega|>m$, the term involving the error term~$R_\infty$ squared
is integrable over~$r$ and can again be
absorbed into the function~$h(m,m')$. The term involving one exponential factor and one
error term~$R_\infty$ is not integrable in the Lebesgue sense, but it exists as
an improper Riemann integral, giving a bounded function.
Hence it can also be absorbed into the function~$h(m,m')$.
Therefore, it remains to consider the $r$-integral for the leading term in~\eqref{eq:3s}.
Transforming for convenience to the Regge-Wheeler coordinate, we obtain
\begin{align*}
R &:= \int_{r_1}^\infty \frac{r}{\sqrt{\Delta}} \: 
\Big\la X^{\omega}_{m,a}(r), \begin{pmatrix} 0 & 1 \\ 1 & 0 \end{pmatrix}
X^{\omega}_{m,a'}(r) \Big\ra_{\C^2} \: e^{-\varepsilon r} \:dr \\
&\,= \int_0^\infty 
\Big\la A^\omega_m \begin{pmatrix} e^{-i \Phi^\omega_m(u)} \:f_{\infty,\: m,a}^{\omega +} \\
e^{i \Phi^\omega_m(u)} \:f_{\infty,\:m,a}^{\omega -} \end{pmatrix}, \begin{pmatrix} 0 & 1 \\ 1 & 0 \end{pmatrix}
A^\omega_{m'} \begin{pmatrix} e^{-i \Phi^\omega_{m'}(u)} \:f_{\infty,\:m', a'}^{\omega +} \\
e^{i \Phi^\omega_{m'}(u)} \:f_{\infty,\:m', a'}^{\omega -} \end{pmatrix} \Big\ra_{\C^2} 
e^{-\varepsilon u} \:du \\
&\quad\: + h(m,m')
\end{align*}
(here it suffices to integrate over~$[0, \infty)$ with the integration measure~$du$ because
the error can again be absorbed into the function~$h(m,m')$).
Using the formula for the phase~$\Phi^\omega_m$ in~\eqref{eq:3u},
one finds that the contributions involving the sum of the phases
$\pm (\Phi^\omega_m(u) + \Phi^\omega_{m'}(u))$ are finite due to the oscillations
even in the limit~$\varepsilon \searrow 0$ and can thus be absorbed into the function~$h(m,m')$.
We thus obtain
\begin{align*}
R &= \int_0^\infty \Big( \big(A^\omega_m \big)^* \begin{pmatrix} 0 & 1 \\ 1 & 0 \end{pmatrix} A^\omega_{m'}
\Big)^1_1\; \overline{f_{\infty,\: m,a}^{\omega +}} \,f_{\infty,\:m', a'}^{\omega +}\;
e^{i \Phi^\omega_m(u) - i  \Phi^\omega_{m'}(u) - \varepsilon u}\: du \\
&\quad\: +\int_0^\infty \Big( \big(A^\omega_m \big)^* \begin{pmatrix} 0 & 1 \\ 1 & 0 \end{pmatrix} A^\omega_{m'}
\Big)^2_2\; \overline{f_{\infty,\: m,a}^{\omega -}} \,f_{\infty,\:m', a'}^{\omega -}\;
e^{-i \Phi^\omega_m(u) + i  \Phi^\omega_{m'}(u) - \varepsilon u}\: du \\
&\quad\: + h(m,m')
\end{align*}
with a new function~$h \in L^\infty(I \times I)$.
Now a straightforward computation using the explicit form of the matrix~$A$ in~\eqref{eq:3v}
and the phase in~\eqref{eq:3u} yields
\begin{align*}
R &= -\epsilon(\omega)\: \frac{m}{\omega^2-m^2} \: \Bigg(
\frac{\overline{f_{\infty,\: m,a}^{\omega +}} \,f_{\infty,\:m', a'}^{\omega +}}{
i \epsilon(\omega) \Big( \sqrt{\omega^2-m^2} - \sqrt{\omega^2 - m'^2} \Big) - \varepsilon} \\
&\qquad\qquad\qquad\quad\;\;\:
+ \frac{\overline{f_{\infty,\: m,a}^{\omega -}} \,f_{\infty,\:m', a'}^{\omega -}}{
i \epsilon(\omega) \Big( -\sqrt{\omega^2-m^2} + \sqrt{\omega^2 - m'^2} \Big) - \varepsilon}
\Bigg) + h(m,m')
\end{align*}
(again with a new function~$h \in L^\infty(I \times I)$). Linearizing near~$m=m'$ gives
the distributional equation
\begin{align}
\lim_{\varepsilon \searrow 0} R &= -\lim_{\varepsilon \searrow 0} \bigg(
\frac{\overline{f_{\infty,\: m,a}^{\omega +}} \,f_{\infty,\:m', a'}^{\omega +}}
{i (m-m') - \epsilon(\omega)\: \varepsilon}
+ \frac{\overline{f_{\infty,\: m,a}^{\omega -}} \,f_{\infty,\:m', a'}^{\omega -}}
{-i (m-m') - \epsilon(\omega)\: \varepsilon} \bigg) + h(m,m') \label{fpole} \\
&= -\pi\, \epsilon(\omega)\, \delta(m-m')\: \Big( \overline{f_{\infty,\: m,a}^{\omega +}} \,f_{\infty,\:m, a'}^{\omega +} 
+ \overline{f_{\infty,\: m,a}^{\omega -}} \,f_{\infty,\:m, a'}^{\omega -}  \Big) \\
&\quad\:+ i\, \frac{\text{PP}}{m-m'}\:\Big( \overline{f_{\infty,\: m,a}^{\omega +}} \,f_{\infty,\:m', a'}^{\omega +} 
- \overline{f_{\infty,\: m,a}^{\omega -}} \,f_{\infty,\:m', a'}^{\omega -} \Big) \label{fcombi}
+ h(m,m') \:.
\end{align}
The combination of transmission coefficients in~\eqref{fcombi} can be simplified using the
relation~\eqref{fpseudo} to obtain
\[ \overline{f_{\infty,\: m,a}^{\omega +}} \,f_{\infty,\:m', a'}^{\omega +} 
- \overline{f_{\infty,\: m,a}^{\omega -}} \,f_{\infty,\:m', a'}^{\omega -} 
= \O \big(m-m' \big) + \delta_{a,a'}\: \times \left\{ \begin{array}{cl} 1 & \text{if~$a=1$} \\
-1 & \text{if~$a=2$} \:. \end{array} \right.  \]
Using these formulas in~\eqref{ppex} gives the result.
\QED

\Proof[Proof of Theorem~\ref{thmstiprep}]
The formula comes about by combining the results of the previous two lemmas:
While Lemma~\ref{lemmapint} determines the contribution for~$m \neq m'$,
Lemma~\ref{lemmasing} tells us about the singular behavior at~$m=m'$.

In order to compute the contribution for~$m \neq m'$, we assume that the
mass supports of~$\hat{\psi}_{.,a}(\omega)$
and~$\hat{\phi}_{.,a'}(\omega)$ are disjoint. Then, according to~\eqref{stipsep},
\begin{align*}
&\bra \p \psi \,|\, \p T \phi \ket - \bra \p T \psi \,|\, \p \phi \ket = -4 \pi \int_{r_1}^\infty dr\: \frac{r}{\sqrt{\Delta}} 
\int_I dm \int_I dm' \int_{-\infty}^\infty d\omega \notag \\
&\qquad \times \sum_{a, a'=1}^2 
(m'-m) \:\overline{\hat{\psi}_{m,a}(\omega)}\: \hat{\phi}_{m',a'}(\omega)\: 
\Big\la X^{k \omega n}_{m,a}(r), \begin{pmatrix} 0 & 1 \\ 1 & 0 \end{pmatrix}
X^{k \omega' n}_{m',a'}(r) \Big\ra_{\C^2} \:.
\end{align*}
Since~$\hat{\psi}_{m,a}$ and~$\hat{\phi}_{m',a'}$ may be multiplied by arbitrary test functions in~$m$
and~$m'$, respectively, comparing again with~\eqref{stipsep} and the formula of Lemma~\ref{lemmapint},
we obtain
\begin{align*}
\bra \p \psi \,|\, \p \phi \ket &= -4 \pi i \int_I dm \int_I dm' \:\frac{1}{m-m'} \\
&\qquad\quad\times
\int_{-\infty}^\infty \sum_{a,a'=1}^2 \overline{\hat{\psi}_{m,a}(\omega)}\: \hat{\phi}_{m',a'}(\omega) \;
\la f^{\omega}_{0,\: m,a} , \begin{pmatrix} 1 & 0 \\ 0 & -1 \end{pmatrix} f^{\omega}_{0,\: m',a'} \ra_{\C^2}\:
d\omega \:.
\end{align*}
We have thus derived the contribution for~$m \neq m'$ in the formula of the theorem.
Moreover, the summands \eqref{print1} and~\eqref{print22} also appear in the above formula. 

It remains to show that in the case~$|\omega|<m$, the integrals involving the principal part
can be combined with the function~$h$ in~\eqref{print3}.
In other words, our task is to show that the function
\beq \label{integrand}
\chi_{[-m,m]}(\omega) \: 
\frac{1}{m-m'} \;\overline{\hat{\psi}_{m,1}(\omega)}\: \hat{\phi}_{m',1}(\omega) \;
\la f^{\omega}_{0,\: m,1} , \begin{pmatrix} 1 & 0 \\ 0 & -1 \end{pmatrix} f^{\omega}_{0,\: m',1} \ra_{\C^2}
\eeq
is bounded (meaning that the corresponding integrals are well-defined even
in the Lebesgue sense without taking a principal value). To this end, we note that in the case~$|\omega|<m$,
the fundamental solution~$X^\omega_{m,1}(r)$ tends to zero at spatial infinity.
Therefore, the differential equation~\eqref{Xnormdiff} implies that
\beq \label{finf0}
\la f^{\omega}_{0,\: m,1} , \begin{pmatrix} 1 & 0 \\ 0 & -1 \end{pmatrix} f^{\omega}_{0,\: m,1} \ra_{\C^2} 
= 0 \:.
\eeq
As a consequence, the corresponding inner product in~\eqref{integrand} is of the order~$\O(m-m')$,
giving the result.

\QED

\subsection{Interpretation of the Mass Decomposition} \label{secinterpret}
We now explain how the different terms in the mass decomposition
of Theorem~\ref{thmstiprep} come about and discuss their significance.
The structure of the mass decomposition becomes clearer if we write it
in the form~\eqref{massdecomp1} and~\eqref{massdecomp2}, where we set
\beq \label{stipm}
(\psi_m \,|\, \Sig_m \,\phi_m)_m := 
4 \pi^2 \int_{\R \setminus [-m,m]} \!\!\!\!\!\!\!\!\!\!\!\epsilon(\omega)\: d\omega \sum_{a, a'=1}^2
\overline{\hat{\psi}_{m,a}(\omega)}\: \hat{\phi}_{m,a'}(\omega) \:
\la f_{\infty,\: m,a}^{\omega}, \,f_{\infty,\:m, a'}^{\omega} \ra_{\C^2} \:,
\eeq
and~${\mathfrak{B}}(\psi_m, \phi_{m'})$ stands for the integrands of the principal
value integrals in Theorem~\ref{thmstiprep}.
The relation~\eqref{stipm} will serve as the definition of the fermionic signature
operator~$\Sig_m$ (see Section~\ref{secSig}); for the moment, the left side
of~\eqref{stipm} merely is a convenient abbreviation.

The term~\eqref{massdecomp1} involves a single mass integral.
Intuitively speaking, this term can be understood from the fact that the {\em{mass
oscillations}} give a factor~$\delta(m-m')$, making it possible to carry out one of the
mass integrals. Consequently, the contribution~\eqref{massdecomp1} tells us about
the behavior of the Dirac wave functions in the asymptotic end for large times,
as shown in Figure~\ref{penrose} in a conformal diagram.
\begin{figure}
\caption{Propagation of Dirac waves in a conformal diagram.}
\label{penrose}
\end{figure}%
In this diagram, a spatially compact solution is shown (the gray region denotes the support).
Clearly, only the diamond up to the horizons~${\mathcal{H}}^\pm$ belongs to
our space-time.
The mass oscillations come into play at lightlike and timelike infinity.
The term~\eqref{massdecomp2}, on the other hand, arises in our analysis
as the boundary terms when integrating the Dirac operator by parts
(see the proof of Lemma~\ref{lemmapint}). With this in mind, these contributions
can be understood as {\em{boundary terms}} on the event horizon (as shown in Figure~\ref{penrose}).
Usually, such boundary terms describe the flux of Dirac currents.
In our setting, the situation is a bit more involved because the
masses of the wave functions~$\psi_m$ and~$\phi_{m'}$ are
in general different. But at least in the limit when the masses coincide,
\[ \lim_{m' \rightarrow m} {\mathfrak{B}}(\psi_m, \phi_{m'}) \:, \]
the integrand of the double mass integral goes over to the probability flux
of the Dirac current through the event horizon. This can be made more
precise by introducing the so-called {\em{fermionic flux operator}}
similar to~\eqref{stipm} by (for details see Section~\ref{secB})
\beq \label{Bdef}
(\psi_m \,|\, \B_m \,\phi_m)_m := \lim_{m' \rightarrow m} {\mathfrak{B}}(\psi_m, \phi_{m'}) \:.
\eeq

An interesting point is that the boundary terms and the terms arising from the
mass oscillations are not independent of each other, but they come from a joint
pole structure (as one sees best in~\eqref{fpole}).
This phenomenon can be understood from the unitarity of the
time evolution, which gives connections between the behavior of the wave
on the event horizon and in the asymptotically flat end.

We finally remark that a similar connection between boundary terms and
double mass integrals involving a principal value is found
in cosmological De Sitter space-time~\cite{desitter}.

\section{The Fermionic Signature Operator} \label{secSig}
The next question is what information on the solution space~$\H_m$ for fixed~$m$
can be extracted from the mass decomposition of Theorem~\ref{thmstiprep}.
One method is to analyze the integrand in the first line; this will be done in this section.
Another method is to analyze the integrand of the double integrals
in the limiting case~$m' \rightarrow m$; this will be explained in
Section~\ref{secB} below.

The main result of this section is to show that~\eqref{stipm}
uniquely defines the {\em{fermionic signature operator}}~$\Sig_m$
as an operator with the following properties:
\begin{Thm} \label{thmsig} Restricting attention to one angular momentum mode, 
for all~$\psi_m \in \H^\infty_m$ the fermionic
signature operator defined by~\eqref{stipm} has the alternative representations
\begin{align}
&\Sig_m \psi_m = \frac{1}{2 \pi^2} \int_{\R \setminus [-m,m]} \frac{\epsilon(\omega)}
{\|f_{\infty,\:m, 1}^{\omega}\|_{\C^2}^2+1} \:\bigg( \sum_{a=1}^2 \label{res1}
\Psi^{\omega}_{m,a} \:(\Psi^\omega_{m,a} \,|\, \psi_m  )_m \bigg) d\omega \\
&\big(\widehat{\Sig_m \psi_m}\big)_{m,a}(\omega) = 
\frac{\epsilon(\omega)\:\chi_{\R \setminus [-m,m]} }{\|f_{\infty,\:m, 1}^{\omega}\|_{\C^2}^2 + 1}\:
\sum_{b=1}^2\: \big(T(\omega)^{-1}\big)^{ab} \: \hat{\psi}_{m,b}(\omega) \:. \label{res2}
\end{align}
The fermionic signature operator is a bounded symmetric operator on~$\H_m$ with
\beq \label{Sigbound}
\|\Sig_m\| \leq 2 \:.
\eeq
Moreover, it commutes with the Dirac Hamiltonian~$H$. It has the spectral representation
\beq \label{Sigspecrep}
\Sig_m = \int_{\R \setminus [-m,m]} \Sig_m(\omega)\: dE_\omega \:,
\eeq
where~$E$ is the spectral measure of the Hamiltonian (see~\eqref{Edef})
with operators~$\Sig_m(\omega)$ having the eigenvalues~\eqref{mupm}.
\end{Thm}

The remainder of this section is devoted to the proof of this theorem.
We begin with the representation~\eqref{lemmaprintrep},
\[ ( \psi_m \,|\, \Sig_m \phi_m)_ m = 
2 \pi^2 \int_\R d\omega \sum_{a,b=1}^2 \big(T(\omega)^{-1}\big)^{ab}\;
\overline{\hat{\psi}_{m,a}(\omega)} \:(\widehat{\Sig_m \phi_m})_{m,b}(\omega) \:. \]
Comparing with~\eqref{stipm} gives
\begin{align*}
&\frac{1}{2}\int_\R d\omega \sum_{a,b=1}^2 \big(T(\omega)^{-1}\big)^{ab}\;
\overline{\hat{\psi}_{m,a}(\omega)} \:(\widehat{\Sig_m \phi_m})_{m,b}(\omega) \\
&=\int_{\R \setminus [-m,m]} \!\!\!\!\!\!\!\!\!\epsilon(\omega)\: d\omega \sum_{a, a'=1}^2
\overline{\hat{\psi}_{m,a}(\omega)}\: \hat{\phi}_{m,a'}(\omega) \:
\la f_{\infty,\: m,a}^{\omega}, \,f_{\infty,\:m, a'}^{\omega} \ra_{\C^2} \\
&=\int_{\R \setminus [-m,m]} \!\!\!\!\!\!\!\!\!\epsilon(\omega)\: d\omega \sum_{a, b,c, a'=1}^2
\big(T(\omega)^{-1}\big)^{ab}\:
\overline{\hat{\psi}_{m,a}(\omega)}\:  \Big( t^\omega_{bc} \:
\la f_{\infty,\: m,c}^{\omega}, \,f_{\infty,\:m, a'}^{\omega} \ra_{\C^2}\:\hat{\phi}_{m,a'}(\omega) \Big)\:.
\end{align*}
Hence
\[ (\widehat{\Sig_m \phi_m})_{m,b}(\omega) = 
2\,\epsilon(\omega)\: \chi_{\R \setminus [-m,m]}(\omega) \sum_{c, a'=1}^2
t^\omega_{bc} \:
\la f_{\infty,\: m,c}^{\omega}, \,f_{\infty,\:m, a'}^{\omega} \ra_{\C^2}\:\hat{\phi}_{m,a'}(\omega) \:. \]
Using~\eqref{superposition} and~\eqref{hatpsi}, we obtain
\begin{align*}
&\big( \Sig_m \phi_m \big)(r,\vartheta,\varphi) = \int_\R \sum_{b=1}^2
(\widehat{\Sig_m \phi_m})_{m,b}(\omega)\: \Psi^{\omega}_{m,b}(r,\vartheta,\varphi) \: d\omega \\
&= 2 \int_{\R \setminus [-m,m]} \!\!\!\!\!\!\!\!\!\epsilon(\omega)\: d\omega \sum_{b, c, a'=1}^2
t^\omega_{bc} \: \la f_{\infty,\: m,c}^{\omega}, \,f_{\infty,\:m, a'}^{\omega} \ra_{\C^2}\:\hat{\phi}_{m,a'}(\omega)
\: \Psi^{\omega}_{m,b}(r,\vartheta,\varphi) \\
&= \frac{1}{\pi^2} \int_{\R \setminus [-m,m]} \!\!\!\!\!\!\!\!\!\epsilon(\omega)\: d\omega \sum_{b, c, a',d=1}^2
t^\omega_{bc} \: \la f_{\infty,\: m,c}^{\omega}, \,f_{\infty,\:m, a'}^{\omega} \ra_{\C^2}\:
t^\omega_{a' d} \:(\Psi^\omega_{m,d} \,|\, \phi_m )_m
\: \Psi^{\omega}_{m,b}(r,\vartheta,\varphi) \:.
\end{align*}
We conclude that
\beq \label{Sig0}
\Sig_m = \frac{1}{\pi^2} \int_{\R \setminus [-m,m]} \!\!\!\!\!\!\!\!\!\epsilon(\omega)\: d\omega \sum_{a, b, c, d=1}^2
\Psi^{\omega}_{m,a} \: t^\omega_{ab} \: \la f_{\infty,\: m,b}^{\omega}, \,f_{\infty,\:m, c}^{\omega} \ra_{\C^2}\:
t^\omega_{c d} \:(\Psi^\omega_{m,d} \,|\, .  )_m \:.
\eeq
The combination of transmission coefficients and matrix elements~$t^\omega_{ab}$ appearing here
is computed in the next lemma.

\begin{Lemma} \label{lemmatab}
For all~$|\omega|>m$,
\beq \label{tabrel}
\sum_{b, c=1}^2 t^\omega_{ab} \: \la f_{\infty,\: m,b}^{\omega}, \,f_{\infty,\:m, c}^{\omega} \ra_{\C^2}\: t^\omega_{c d} 
= \frac{\delta_{ad}}{2 \,\big(1 + \|f_{\infty,\:m, a}^{\omega} \|_{\C^2}^2\big)} \:.
\eeq
\end{Lemma} 
\Proof
We make use of the explicit formulas for the coefficients~$t^\omega_{ab}$ as 
derived in~\cite[Theorem~3.6]{tkerr} and~\cite[Lemma~6.1]{decay}.
We first recall these results and formulate them in a way most convenient for us.
In~\cite{tkerr} the integral representation is obtained by first analyzing the
system with Dirichlet boundary conditions at~$u_2 \in \R$ and then taking the
limit~$u_2 \rightarrow \infty$. Considering a linear combination of the radial fundamental solutions
\[ X(u) = c_1\: X_1(u) + c_2\: X_2(u) \:, \]
the Dirichlet boundary conditions take the form~$X_+(u_2)=X_-(u_2)$.
Evaluating these conditions asymptotically as~$u \rightarrow \infty$
with the help of~\eqref{eq:3s} and keeping in mind
that the normalization at the event horizon implies that~$|c_1|^2 + |c_2|^2=1$, one finds
\[ c_1 = \frac{t_1}{\sqrt{|t_1|^2+|t_2|^2}}\:,\qquad c_2 = \frac{t_2}{\sqrt{|t_1|^2+|t_2|^2}} \]
with
\beq \label{t12def}
t_1(\alpha) = f_{\infty\: 2}^+ \:e^{-i \alpha} - f_{\infty\: 2}^- \:e^{i \alpha}
\:,\qquad t_2(\alpha) = -f_{\infty\:1}^+ \:e^{-i \alpha} + f_{\infty\:1}^- \:e^{i \alpha}
\eeq
and~$\alpha=\Phi^\omega_m(u)$.
The coefficients~$t^\omega_{ab}$ are obtained by taking the product~$c_a \overline{c_b}$
and integrating over~$\alpha$ (see~\cite[eq.~(3.46)]{tkerr}),
\beq
t^\omega_{ab} = \frac{1}{2 \pi} \:\int_0^{2 \pi} \frac{t_a \:\overline{t_b}}{|t_1|^2 + |t_2|^2} \:d\alpha \:. \label{eq:3H2}
\eeq
In~\cite{decay} more detailed formulas for~$t^\omega_{ab}$ were derived using specific properties
of the radial equation which become most apparent in the reformulation as the so-called
{\em{planar equation}} (see~\cite[Section~4]{decay}). For our purposes, it suffices
and is more convenient to incorporate the additional properties of the radial equation
by employing the methods and results of Lemma~\ref{lemmapseudo}.

Again in the parametrization~\eqref{fparam}, the scalar products in~\eqref{tabrel} become
\beq \label{prods}
\begin{split}
\la f_{\infty,\: m,1}^{\omega}, \,f_{\infty,\:m, 1}^{\omega} \ra_{\C^2} &= 
\la f_{\infty,\: m,2}^{\omega}, \,f_{\infty,\:m, 2}^{\omega} \ra_{\C^2} = \cosh (2 \vartheta) \\
\la f_{\infty,\: m,1}^{\omega}, \,f_{\infty,\:m, 2}^{\omega} \ra_{\C^2} &= e^{i \delta}\: \sinh (2 \vartheta) \:.
\end{split}
\eeq
Moreover, using this parametrization in~\eqref{t12def}, a short computation shows that
\[ |t_1|^2 = |t_2|^2 \:. \]
Applying this relation in~\eqref{eq:3H2}, we immediately find
\beq \label{t11t22}
t^\omega_{11} = t^\omega_{22} = \frac{1}{2}
\eeq
and
\begin{align*}
t^\omega_{12} &= \frac{1}{4 \pi} \:\int_0^{2 \pi} \frac{t_1 \:\overline{t_2}}{|t_2|^2} \:d\alpha
= \frac{1}{4 \pi} \:\int_0^{2 \pi} \frac{t_1}{t_2} \:d\alpha \\
&= \frac{1}{2 \pi} \:\int_0^{\pi} \frac{f_{\infty\: 2}^+ - f_{\infty\: 2}^- \:e^{2 i \alpha}}
{-f_{\infty\:1}^+ + f_{\infty\:1}^- \:e^{2 i \alpha}} \:d\alpha \:.
\end{align*}
Introducing~$z=e^{2 i \alpha}$ as the new integration variable, we obtain the contour integral
\begin{align*}
t^\omega_{12} &= \frac{1}{2 \pi} \:\ointctrclockwise_{\partial B_1(0)} 
\frac{f_{\infty\: 2}^+ - f_{\infty\: 2}^- \:z}
{-f_{\infty\:1}^+ + f_{\infty\:1}^- \:z} \:\Big( -\frac{i}{2}\: \frac{dz}{z} \Big) \:.
\end{align*}
The integrand has poles at~$z=0$ and
\[ z = \frac{f_{\infty\:1}^+}{f_{\infty\:1}^-} \overset{\eqref{fparam}}{=} e^{i (\beta-\gamma)}\: \coth \vartheta \:. \]
Since the last pole lies outside our integration contour, we only need to take into account the contour at~$z=0$.
We thus obtain
\beq \label{t12}
t^\omega_{12} = -\frac{1}{2} \:\frac{f_{\infty\: 2}^+}{f_{\infty\:1}^+} = -\frac{1}{2}\: e^{i \delta}\: \tanh \vartheta \:.
\eeq
Finally, the coefficient~$t^\omega_{21}$ is obtained by complex conjugation,
\beq \label{t21}
t^\omega_{21} = \overline{t^\omega_{12}} = -\frac{1}{2}\: e^{-i \delta}\: \tanh \vartheta \:.
\eeq

Combining the identities~\eqref{t11t22}, \eqref{t12} and~\eqref{t21} with~\eqref{prods},
a straightforward computation yields
\[ \sum_{b, c=1}^2 t^\omega_{ab} \: \la f_{\infty,\: m,b}^{\omega}, \,f_{\infty,\:m, c}^{\omega} \ra_{\C^2}\:
t^\omega_{c d} = \frac{\delta_{ad}}{4}\: \frac{1}{\cosh^2 \vartheta} \:. \]
Rewriting the factor~$\cosh^2 \vartheta$ as the absolute square
of the vectors in~\eqref{fparam}
\beq \label{cosh2}
\cosh^2 \vartheta = \frac{1}{2} \big( \cosh^2 \vartheta + \sinh^2 \vartheta +1 \big)
= \frac{1}{2} \:\Big( \|f^\omega_{\infty,\:m,a}\|_{\C^2}^2 +1 \Big)
\eeq
concludes the proof.
\QED

\Proof[Proof of Theorem~\ref{thmsig}]
The representation~\eqref{res1} follows immediately by using the identity
of Lemma~\ref{lemmatab} in~\eqref{Sig0}.
Applying~\eqref{superposition} and~\eqref{hatpsi} gives~\eqref{res2}.

According to~\eqref{res2}, the fermionic signature operator is a multiplication operator in~$\omega$.
This implies that it commutes with the Hamiltonian and can be represented in the form~\eqref{Sigspecrep}.
Moreover, the eigenvalues~$\mu_\pm$ in~\eqref{mupm} are the eigenvalues
of the matrix in~\eqref{res2}. In order to compute them, we
again work in the parametrization~\eqref{fparam}. Then
\[ T(\omega) = \frac{1}{2} \begin{pmatrix} 1 & -e^{i \delta}\: \tanh \vartheta \\
-e^{-i \delta}\: \tanh \vartheta & 1 \end{pmatrix} \:. \]
This matrix has the eigenvalues
\[ \nu_\pm = \frac{1}{2}\: (1 \mp \tanh \vartheta )\:. \]
Thus the matrix in~\eqref{res2} has the eigenvalues
\[ \mu_\pm = \frac{\epsilon(\omega)}{2\, \cosh^2 \vartheta}\: \frac{1}{\nu_\pm}
= \epsilon(\omega) \pm \tanh \vartheta \:. \]
Finally, we express the hyperbolic tangent in terms of the norm of the vectors in~\eqref{fparam},
\[ \tanh^2 \vartheta = \frac{\cosh^2 \vartheta + \sinh^2 \vartheta - 1}{\cosh^2 \vartheta + \sinh^2 \vartheta +1} 
= \frac{\|f^\omega_{\infty,\:m,a}\|_{\C^2}^2 - 1}{\|f^\omega_{\infty,\:m,a}\|_{\C^2}^2 +1} \:. \]
This concludes the proof.
\QED

\section{The Fermionic Flux Operator} \label{secB}
In this section we shall analyze how one can
extract information on the solution space~$\H_m$ for fixed~$m$
from the double integral in Theorem~\ref{thmstiprep}.
For convenience, we again write this double integral in the form~\eqref{massdecomp2}.
In Section~\ref{secinterpret} we already mentioned the method
of representing the integrand~${\mathfrak{B}}(\psi_m, \phi_{m'})$
in the limit~$m' \rightarrow m$ in terms of the scalar product on~$\H_m$,
giving rise to the so-called {\em{fermionic flux operator}}~$\B_m$ in~\eqref{Bdef}.
Before entering the details of this construction, we point out that this operator is the {\em{only}}
operator on~$\H_m$ which can be constructed from~\eqref{massdecomp2}.
Indeed, a more general idea would be to expand to higher order in the masses
before taking the limit~$m' \rightarrow m$,
\beq \label{higher}
\lim_{m' \rightarrow m} 
\frac{d^p}{dm^p} \frac{d^q}{dm'^q} {\mathfrak{B}}(\psi_m, \phi_{m'}) \qquad \text{with~$p+q>0$}\:.
\eeq
However, these bilinear forms depend on
how the solutions $\phi_m, \psi_m \in \H_m$ are extended to families of
solutions described by the mass parameter in~$I$.
For this reason, the bilinear forms~\eqref{higher} do not give rise to well-defined
operators on~$\H_m$.

\begin{Thm} \label{thmB} Restricting attention to one angular momentum mode, 
for all~$\psi_m \in \H^\infty_m$ the fermionic
flux operator defined by~\eqref{Bdef} has the alternative representations
\begin{align}
&\B_m \psi_m = -\frac{1}{2 \pi^2} \int_{\R \setminus [-m,m]} \frac{1}
{\|f_{\infty,\:m, 1}^{\omega}\|_{\C^2}^2+1} \:\bigg( \sum_{a=1}^2
s_a\: \Psi^{\omega}_{m,a} \:(\Psi^\omega_{m,a} \,|\, \psi_m  )_m \bigg) d\omega \label{Bres1} \\
&\big(\widehat{\B_m \psi_m}\big)_{m,a}(\omega) =
-\frac{\chi_{\R \setminus [-m,m]} }{\|f_{\infty,\:m, 1}^{\omega}\|_{\C^2}^2 + 1}\:
\sum_{b=1}^2 s_a\: \big(T(\omega)^{-1}\big)^{ab} \: \hat{\psi}_{m,b}(\omega) \:, \label{Bres2}
\end{align}
where
\beq \label{sdef}
s_1=1 \qquad \text{and} \qquad s_2=-1 \:.
\eeq
The fermionic flux operator is a bounded symmetric operator on~$\H_m$ with
\[ \|\B_m\| \leq 1 \:. \]
It commutes with the Dirac Hamiltonian~$H$. It has the spectral representation
\[ \B_m = \int_{\R \setminus [-m,m]} \B_m(\omega)\: dE_\omega \:, \]
where~$E$ is the spectral measure of the Hamiltonian (see~\eqref{Edef})
with operators~$\B_m(\omega)$ having the eigenvalues
\[ \nu_\pm(\omega) = \pm \sqrt{\frac{2}{\|f^{k\omega n}_{\infty,\:m,1} \|^2_{\C^2} + 1}}\:. \]
\end{Thm} \noindent
Comparing with Theorem~\ref{thmsig}, one sees that the spectral decompositions
of the fermionic flux operator and the fermionic signature operator are quite different.
Indeed, the sign of the eigenvalues of~$\B_m$ does not depend on the sign of~$\omega$.
Instead, the negative spectral subspace of~$\B_m$ describes the Dirac waves
which ``enter the black hole,'' whereas the positive spectral subspace corresponds to
Dirac waves which ``emerge from the white hole.''
\Proof[Proof of Theorem~\ref{thmB}]
Comparing the formula in Theorem~\ref{thmstiprep} with~\eqref{massdecomp1} and~\eqref{massdecomp2},
we obtain
\begin{align*}
(\psi_m | \B_m \phi_m)_m &= 
-4 \pi^2 \int_{-\infty}^\infty \sum_{a,a'=1}^2
\overline{\hat{\psi}_{m,a}(\omega)}\: \hat{\phi}_{m,a'}(\omega) \;
\la f^{\omega}_{0,\: m,a} , \begin{pmatrix} 1 & 0 \\ 0 & -1 \end{pmatrix} f^{\omega}_{0,\: m,a'} \ra_{\C^2}\:
d\omega \:.
\end{align*}
The inner product on the right can be further simplified: In the case~$|\omega|<m$,
according to~\eqref{decayfund} we only need to consider the contributions for~$a=a'=1$.
These contributions vanish according to~\eqref{finf0}.
Therefore, we do not get a contribution if~$|\omega|<m$.

In the remaining case~$|\omega|>m$, we can use the identity~\eqref{fevent}. We thus obtain
\begin{align*}
(\psi_m | \B_m \phi_m)_m &= 
-4 \pi^2 \int_{\R \setminus [-m,m]} \Big(
\overline{\hat{\psi}_{m,1}(\omega)}\: \hat{\phi}_{m,1}(\omega)
- \overline{\hat{\psi}_{m,2}(\omega)}\: \hat{\phi}_{m,2}(\omega) \Big) \:d\omega \:.
\end{align*}
Introducing the coefficients~$s_a$ by~\eqref{sdef}, we can write
this formula as
\begin{align*}
(\psi_m | \B_m \phi_m)_m &= -4 \pi^2 \int_{\R \setminus [-m,m]} \: \sum_{a=1}^2 s_a \:
\overline{\hat{\psi}_{m,a}(\omega)}\: \hat{\phi}_{m,a}(\omega) \:d\omega \\
&= -4 \pi^2 \int_{\R \setminus [-m,m]} \: \sum_{a,b,c=1}^2 
\big( T(\omega)^{-1} \big)^{ab} \:
\overline{\hat{\psi}_{m,a}(\omega)}\: t^\omega_{bc}\: s_c \:\hat{\phi}_{m,c}(\omega) \:d\omega \:.
\end{align*}
Comparing this formula with~\eqref{rep1} gives
\[ (\widehat{\B_m \phi_m})_{m,b} = -2\: \chi_{\R \setminus [-m,m]}(\omega) \sum_{c=1}^2
t^\omega_{bc}\: s_c \:\hat{\phi}_{m,c}(\omega)\:. \]
Using~\eqref{superposition} and~\eqref{hatpsi}, we obtain
\begin{align*}
\big( \B_m \phi_m \big)(r,\vartheta,\varphi) &= \int_\R \sum_{b=1}^2
(\widehat{\B_m \phi_m})_{m,b}(\omega)\: \Psi^{\omega}_{m,b}(r,\vartheta,\varphi) \: d\omega \\
&= -2 \int_{\R \setminus [-m,m]} \!\!\!\!\!\!\!\!\!d\omega \sum_{b, c=1}^2
t^\omega_{bc}\: s_c \:\hat{\phi}_{m,c}(\omega)\:  \Psi^{\omega}_{m,b}(r,\vartheta,\varphi) \\
&= -\frac{1}{\pi^2} \int_{\R \setminus [-m,m]} \!\!\!\!\!\!\!\!\!d\omega \sum_{b, c, d=1}^2
t^\omega_{bc}\: s_c\:
t^\omega_{c d} \:(\Psi^\omega_{m,d} \,|\, \phi_m )_m
\: \Psi^{\omega}_{m,b}(r,\vartheta,\varphi) \:.
\end{align*}
We conclude that
\[ \B_m = -\frac{1}{\pi^2}  \int_{\R \setminus [-m,m]} \!\!\!\!\!\!\!\!\! d\omega \sum_{a, b, c=1}^2
\Psi^{\omega}_{m,a} \: t^\omega_{ab} \: s_b\:
t^\omega_{bc} \:(\Psi^\omega_{m,c} \,|\, .  )_m \:. \]
A short computation using the explicit formulas for~$t^\omega_{ab}$ as given in~\eqref{t11t22}, \eqref{t12} and~\eqref{t21} yields
\[ \sum_{b=1}^2 t^\omega_{ab} \: s_b\: t^\omega_{bc} = \frac{s_a \,\delta_{ac}}{4\, \cosh^2 \vartheta} \:. \]
Using again~\eqref{cosh2}, we obtain~\eqref{Bres1}.
Applying~\eqref{superposition} and~\eqref{hatpsi} gives~\eqref{Bres2}.

The spectrum of the operator~$\B_m$ is computed similar as in the proof
of Theorem~\ref{thmsig} by diagonalizing the operator in~\eqref{Bres2}.
\QED

\section{Generalized Fermionic Projector States} \label{secFP}
We briefly recall the construction of quasi-free Dirac states as worked out in~\cite{hadamard}.
According to~\eqref{Sigbound}, for any~$m \in I$ the 
fermionic signature operator is a bounded symmetric operator on~$\H_m$.
The {\em{fermionic projector}}~$P$ is introduced as the operator
(for details see~\cite[Section~3]{finite} and~\cite[Section~4.2]{infinite})
\beq \label{Pstar}
P = -\chi_{(-\infty, 0)}(\Sig_m)\, k_m \::\: C^\infty_0(\scrM, S\scrM) \rightarrow \H_m \:,
\eeq
where~$k_m$ is the {\em{causal fundamental solution}} defined as the difference of the
advanced and retarded Green's operators,
\[ k_m := \frac{1}{2 \pi i} \left( s_m^\vee - s_m^\wedge \right) \::\: C^\infty_0(\scrM, S\scrM) \rightarrow 
\H_m^\infty \:. \]
The fermionic projector~$P$ can be written as an
integral operator involving a uniquely determined distributional
{\em{kernel}}~${\mathcal{P}} \in \D'(\scrM \times \scrM)$, i.e.\ (for details see~\cite[Section~3.5]{finite})
\[ 
\bra \phi | P \psi \ket = {\mathcal{P}} \big( \overline{\phi} \otimes \psi \big)
\qquad \text{for all~$\phi, \psi \in C^\infty_0(\scrM, S\scrM)$}\:. \]
Araki's construction in~\cite{araki1970quasifree} yields for any non-negative operator~$W$
on~$\H_m$ a unique quasi-free Dirac state with the property that the two-point distribution
coincides with the integral kernel of the operator~$-W\, k_m$. Applying this construction to the projection
operator~$\chi_{(-\infty, 0)}(\Sig_m)$
gives the so-called {\em{fermionic projector state}} (for details see~\cite[Section~6]{hadamard}).

According to Theorem~\ref{thmsig}, the negative spectral subspace
of the fermionic signature operator coincides with the negative spectral subspace
of the Hamiltonian; more precisely (see also~\eqref{Sigsign})
\[ \chi_{(-\infty, 0)}(\Sig_m) = \chi_{[-\infty, m)}(H) \:. \]
We thus reproduce the frequency splitting for the observer in a rest frame at infinity.
Clearly, this state is Hadamard. This gives the result of Corollary~\ref{corFRV1}.

Applying Araki's construction to the operator~$W=W(\Sig_m)$ with~$W$ a non-negative
Borel function gives the so-called {\em{generalized fermionic projector state}}.
The corresponding two-point distribution is the integral kernel of the operator
\[ 
P_W = -W(\Sig_m)\, k_m \::\: C^\infty_0(\scrM, S\scrM) \rightarrow \H_m \:. \]
In ultrastatic space-times, where the operator~$\Sig_m$ only has the eigenvalues~$\pm1$,
working with~$W(\Sig_m)$ does not give anything new.
This is why the generalized fermionic projector state was first considered in Rindler
space-time~\cite[Section~11]{rindler} (the notion ``generalized fermionic projector state''
was introduced in~\cite[Section~2.5]{sigsymm}).

The basic question is whether the generalized fermionic projector state is a Hadamard state.
We now explain why, for generic~$W$, the generalized fermionic projector
cannot be expected to be a Hadamard state.
To this end, recall that a state is Hadamard if it realizes the frequency splitting
up to smooth contributions. This means in particular that, asymptotically
as~$\omega \rightarrow \pm \infty$, the state should reproduce the frequency splitting
(meaning that~$\lim_{\omega \rightarrow \pm \infty} \mu_s = \pm 1$
for all~$s\in \{+,-\}$).
This condition can be analyzed by looking at the radial equation~\eqref{radial}.
Indeed, as~$\omega \rightarrow \pm \infty$, the potential on the right of this equation
has little effect on the solutions (as could be made precise for example with a WKB analysis),
meaning that the solutions go over asymptotically to plane waves.
As a consequence, the norm~$\|f^{k \omega n}_{\infty, m, 1}\|_{\C^2}$ in~\eqref{mupm} tends to one.
We conclude that
\[ \lim_{\omega \rightarrow \pm \infty} \mu_\pm(\omega) = \epsilon(\omega) \:, \]
implying that, as desired, for large frequencies we recover frequency splitting.
This argument has the caveat that in order to obtain a Hadamard state,
we must recover frequency splitting uniformly in the angular eigenvalue~$\lambda$.
But this uniformity does not hold for the following reason:
Suppose we are given~$\omega$ and a compact interval~$[u_0, u_1]$.
Since the matrix on the right of~\eqref{radial} is Hermitian,
by choosing~$\lambda$ sufficiently large, we can arrange that the fundamental solutions
of the radial equation are exponentially increasing or decreasing on the interval~$[u_0, u_1]$.
This means that, for any fixed~$\omega$,
the norm~$\|f^{k \omega n}_{\infty, m, 1}\|_{\C^2}$ in~\eqref{mupm} can be made
arbitrarily large by increasing~$\lambda$.
Therefore, except in the case when~$W$ is constant on the intervals~$[-2,0)$ and~$(0,2]$
(in which case we get merely a linear combination of the fermionic projector~\eqref{Pstar}
and the operator~$k_m$), the generalized fermionic projector does not
reproduce frequency splitting for large~$\omega$.
As a consequence, for generic~$W$, the generalized fermionic projector state will not be a Hadamard
state.

Clearly, this argument leaves the possibility that one gets a Hadamard state for {\em{specific choices}} of
the function~$W$. If this is the case, the next question would be what this state means physically.
We leave these questions as open problems for the future.

%

\Thanks {{\em{Acknowledgments:}} We would like to thank Niky Kamran for helpful discussions. 
We are grateful to the referees for valuable suggestions. 
\providecommand{\bysame}{\leavevmode\hbox to3em{\hrulefill}\thinspace}
\providecommand{\MR}{\relax\ifhmode\unskip\space\fi MR }
\providecommand{\MRhref}[2]{%
  \href{http://www.ams.org/mathscinet-getitem?mr=#1}{#2}
}
\providecommand{\href}[2]{#2}

\end{document}